\documentclass[conference]{styles/IEEEtran}
\IEEEoverridecommandlockouts
\usepackage{cite}
\usepackage{amsmath,amssymb,amsfonts}
\usepackage{graphicx}
\usepackage{textcomp}
\usepackage{xcolor}
\usepackage{fancyhdr}
\usepackage[hyphens]{url}

\newcommand\hl[1]{#1}

\def\BibTeX{{\rm B\kern-.05em{\sc i\kern-.025em b}\kern-.08em
    T\kern-.1667em\lower.7ex\hbox{E}\kern-.125emX}}

\usepackage[frozencache,cachedir=.]{minted}
\usepackage{subcaption}
\usepackage[font={small,bf}]{caption}
\usepackage{algorithm}
\PassOptionsToPackage{noend}{algpseudocode}
\usepackage{algpseudocode}

\errorcontextlines\maxdimen

\makeatletter
    \newcommand*{\algrule}[1][\algorithmicindent]{\makebox[#1][l]{\hspace*{.5em}\thealgruleextra\vrule height \thealgruleheight depth \thealgruledepth}}%
\newcommand*{\thealgruleextra}{}
\newcommand*{\thealgruleheight}{.75\baselineskip}
\newcommand*{\thealgruledepth}{.25\baselineskip}
\newcommand{\ignore}[1]{}

\newcount\ALG@printindent@tempcnta
\def\ALG@printindent{%
    \ifnum \theALG@nested>0
        \ifx\ALG@text\ALG@x@notext
        \else
            \unskip
            \addvspace{-1pt}
            \ALG@printindent@tempcnta=1
            \loop
                \algrule[\csname ALG@ind@\the\ALG@printindent@tempcnta\endcsname]%
                \advance \ALG@printindent@tempcnta 1
            \ifnum \ALG@printindent@tempcnta<\numexpr\theALG@nested+1\relax
            \repeat
        \fi
    \fi
    }%
\usepackage{etoolbox}
\patchcmd{\ALG@doentity}{\noindent\hskip\ALG@tlm}{\ALG@printindent}{}{\errmessage{failed to patch}}
\makeatother

\newbox\statebox
\newcommand{\myState}[1]{%
    \setbox\statebox=\vbox{#1}%
    \edef\thealgruleheight{\dimexpr \the\ht\statebox+1pt\relax}%
    \edef\thealgruledepth{\dimexpr \the\dp\statebox+1pt\relax}%
    \ifdim\thealgruleheight<.75\baselineskip
        \def\thealgruleheight{\dimexpr .75\baselineskip+1pt\relax}%
    \fi
    \ifdim\thealgruledepth<.25\baselineskip
        \def\thealgruledepth{\dimexpr .25\baselineskip+1pt\relax}%
    \fi
    \State #1%
    \def\thealgruleheight{\dimexpr .75\baselineskip+1pt\relax}%
    \def\thealgruledepth{\dimexpr .25\baselineskip+1pt\relax}%
}

\begin{document}

\title{ Packet Chasing: Spying on Network Packets over a Cache Side-Channel} 
\author{\IEEEauthorblockN{Mohammadkazem Taram \\
{{University of California San Diego}}\\
mtaram@cs.ucsd.edu}
\and
\IEEEauthorblockN{Ashish Venkat \\
{{University of Virginia} }\\
venkat@virginia.edu}
\and
\IEEEauthorblockN{Dean Tullsen\\
{{University of California San Diego}}\\
tullsen@cs.ucsd.edu}
}

\maketitle


\begin{abstract}
This paper presents Packet Chasing, an attack on the network that does not require access to the network, and works regardless of the 
privilege level of the process receiving the packets. 
A spy process can easily probe and discover the exact cache location of each buffer
used by the network driver. 
Even more useful, it can discover the exact sequence in which those buffers are used to receive packets.  
This
then enables packet frequency and packet sizes to be monitored through cache side channels.  
This allows both covert channels
between a sender and a remote spy with no access to the network, as well as direct attacks that can identify,
among other things, the web page access patterns
of a victim on the network.
In addition to identifying the potential attack, this work proposes a software-based short-term mitigation as well as a light-weight,
adaptive, cache partitioning mitigation that blocks the interference of I/O and CPU requests in the last-level cache.

\end{abstract}
\begin{IEEEkeywords}
Side-Channel Attacks, Cache Attacks, DDIO, Packet Processing, Security, Microarchitecture Security
\end{IEEEkeywords}
\section{Introduction}
Modern processors employ increasingly complex microarchitectural techniques that are carefully optimized to deliver high performance.  However, this complexity often breeds security vulnerabilities, as evidenced recently by Meltdown~\cite{meltdown} and Spectre~\cite{spectre}. 
This paper explores the vulnerable side effects of another sophisticated high performance microarchitectural technique -- \emph{Intel\textsuperscript{\textregistered} Data Direct I/O} (DDIO)~\cite{ddio} implemented in most server-grade Intel processors to accelerate network packet processing.  Further, it presents
new high resolution covert and side channel attacks on the network I/O traffic, which while possible without
DDIO, are considerably more effective in the presence of DDIO.

The widespread adoption of multi-gigabit Ethernet and other high-speed network I/O technology such as Infiniband has highlighted the critical importance of processing network packets at high speed in order to sustain this newly available network throughput, and further improve the performance of bandwidth-intensive datacenter workloads.  Consequently, most Intel server-class processors today employ DDIO technology that allows the injection and subsequent processing of network packets directly in the processor's last level cache (LLC), bypassing the traditional DMA (Direct Memory Access) interface.  DDIO is invisible to software, including OS drivers, and is always enabled by default.

The key motivation behind DDIO is the fact that modern server-class processors employ large LLCs ($\sim$20MB in size), thereby allowing the network stack to host hot data structures and network packets in-process completely within the LLC, reducing trips to main memory.  By eliminating redundant memory transfers, DDIO has been shown to provide substantial improvements in I/O bandwidth and overall power consumption~\cite{ddio,li2015architecting,basavaraj2017case,marinos2014network}.  Although Intel restricts allocating more than 10\% of the LLC for DDIO to prevent cache pollution, it neither statically reserves nor dynamically partitions a dedicated portion of the cache for DDIO.

However, despite its good intention to accelerate network packet processing, DDIO has a previously unknown vulnerable side effect that this paper exposes.  On a DDIO host, incoming network packets from a remote client and application data structures from processes on the local host contend for the shared LLC, potentially evicting each other in the event of a cache conflict.  In this paper, we show that such contention provides significant leakage, allowing cache side channel attacks to perform covert communication and/or infer network behavior, with virtually zero access to the network stack.  In particular, we describe a new class of covert- and side-channel cache attacks, called \emph{packet chasing}, that exploit this contention by creating arbitrary conflicts in the LLC using carefully constructed memory access patterns and/or network packet streams.

We further show that the location (in cache) of packet buffers used 
by the network driver, {\em and the order in which they are filled},
are easily discovered by an attacker, greatly minimizing the amount
of probing necessary to follow the sequence of packets being chased.

The \emph{packet chasing}-based covert channel we describe in this paper allows a spy process running covertly alongside a server daemon on the local DDIO host to receive secret messages from a trojan process running on a remote client across the network, by causing deterministic contention in the last-level cache.  We show that such a covert means of communication is feasible, and is achievable at a high bandwidth, despite the fact that the trojan process only sends broadcast packets and that the spy process is completely isolated from the network-facing server daemon (potentially cross-container and cross-VM), and further lacks any access to the network stack.

In addition to the covert channel, we describe a novel \emph{packet chasing} based side-channel attack that leverages a local spy process running alongside (or, within) a web browser. In our experiments, the spy is on the client side alongside of a browser like Mozilla Firefox, enabling it to fingerprint a remote victim's website accesses without having access to the network.  In particular, this attack enables an attacker
to recognize the web activity of the victim based on packet size patterns. {This type of web fingerprinting could be used by an oppressive government to identify accesses to a banned site, or
an attacker could identify members of a secure organization (to then target more directed attacks) simply
by fingerprinting a successful login session.}

{Further, this paper describes \hl{a software-based short-term mitigation, called ring buffer randomization, as well as } a hardware defense mechanism that adaptively partitions the LLC into  I/O and CPU partitions, preventing I/O packets from evicting CPU/adversary cache blocks. The adaptive partitioning defense that we describe in this paper has a performance overhead of less than 2.7\% compared to the vulnerable DDIO baseline.}

The major contributions of this paper are as follows. It shows that (1) with DDIO turned on, the location of the packet buffers for a common network
driver are easily discovered, (2) the size of each packet sent (in cache block increments) is also discoverable, (3) the sequence in which the
discovered buffers are repeatedly accessed can also be deduced, (4) covert channels can be created between a trojan sending packets on the network and a spy on another machine, (5) the sequence/pattern of packet sizes can leak sensitive information across a side channel, such as a trace of 
web access activity, and \hl{(6) a short-term software-only randomization scheme as well as} an adaptive cache partitioning scheme are proposed to defend against the attack with minimal performance overhead.
 \vspace{-2ex}\section{Background and Related Work}

This section provides background on network packet handling, 
DDIO and related network optimizations, network and I/O based attacks, and
cache attacks. 
\vspace{-1ex}\subsection{Journey of a Network Packet}

When an application sends data through the network, it usually sends a stream of data; and it is the responsibility of the transfer layer to break large messages into smaller pieces that the network layer can handle.
The Maximum Transferable Unit (MTU) is the largest contiguous block of data which can be sent across a transmission medium.
For example, the Ethernet MTU is 1500 bytes, which means the largest IP packet (or some other payload) an Ethernet frame can carry is 1500 bytes. 
Adding 26 bytes for the Ethernet header results in a maximum frame of 1526 bytes. 

%
When the NIC driver initializes, it first allocates a buffer for receiving packets and then creates a descriptor which includes the receive buffer size and its physical memory address.
It then adds the receive descriptor to the receive ring (\textit{rx} ring), a circular buffer shared  between the driver and the NIC to store the incoming packets until they can be processed by the driver.
The driver then notifies the NIC that it placed a new descriptor in the \textit{rx} ring. 
The NIC reads the content of the new descriptor and copies the size and the physical address of the buffer into its internal memory.  
At this step, the initialization is done and the NIC is ready to receive packets. 
\begin{figure}
    \centering
    \includegraphics[width=.45\textwidth]{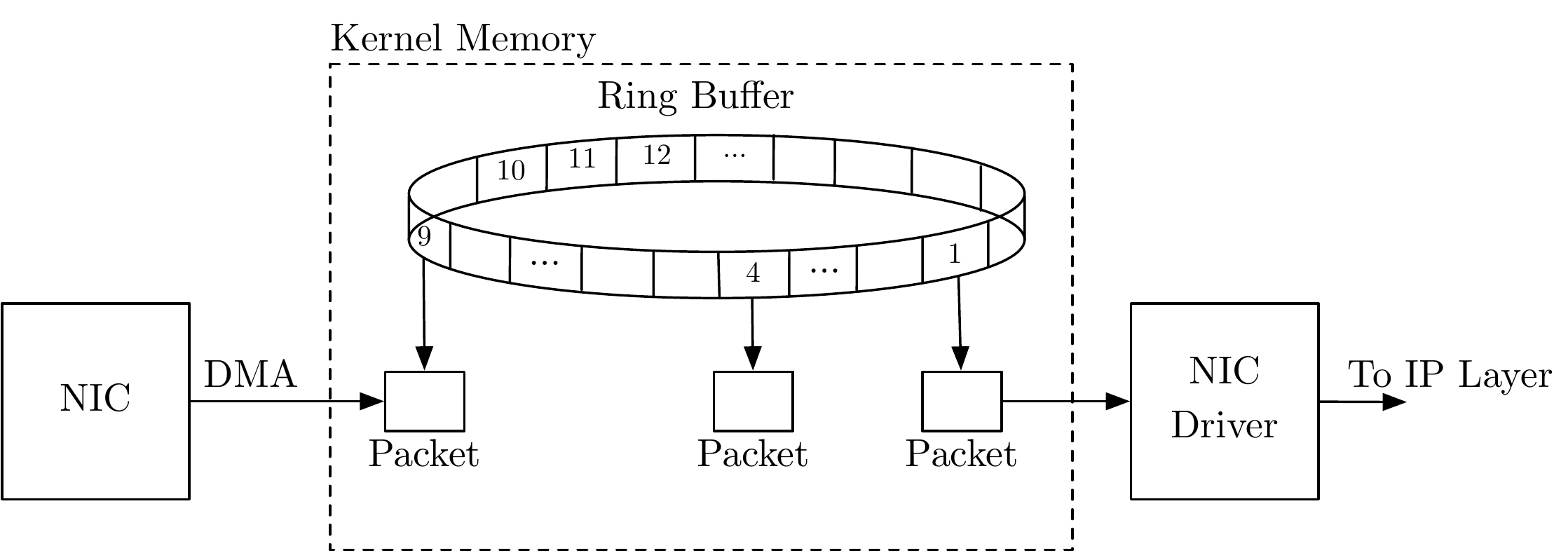}
    \caption{The shared ring buffers (FIFO) between NIC and the device driver.}
    \label{fig:journey}
\end{figure}

As shown in Figure~\ref{fig:journey}, upon receiving incoming packets, the NIC, using Direct Memory Access (DMA), copies packets to the physical addresses provided in the \textit{rx} ring, and then sends an interrupt to notify the driver. 
The driver drains the new packets from the \textit{rx} ring and places each of them in a kernel data structure called a socket buffer (\textit{skb}) to begin their journey through the kernel networking stack up to the application which owns the relevant socket. 
Finally, the driver puts the receive buffer back in the \textit{rx} ring to be used for future packets. 

\subsection{Direct Cache Access and Data Direct I/O}\label{sec:ddio}  
Modern processors and operating systems employ a number of network I/O performance enhancements that address packet processing bottlenecks in the memory subsystem~\cite{ddio, dca}. 
Huggahalli, et al.~\cite{dca}, present Direct Cache Access (DCA), which enables the NIC to provide prefetch hints to the processor's hardware prefetcher.
DCA requires that memory writes go to the host memory and then the processor prefetches the cache lines specified by the memory write.
\ignore{DCA needs to be initialized upon system startup. The I/O device driver configures the DCA by setting up the appropriate DCA target ID for the device. 
The I/O device then encapsulates that information in PCIe Transaction Layer Packet (TLP) headers which then trigger a hardware prefetch to the processor cache~\cite{i350}.
DCA has been part of the Linux Kernel since 2007 and version 2.6.24~\cite{linuxDCA}.
}
The Intel Sandy-Bridge-EP microarchitecture introduced the Data Direct I/O (DDIO) technology~\cite{ddio} which transparently pushes the data from the NIC or other I/O devices directly into the last level cache.
Before DDIO, I/O data was always sent through the main memory; inbound data is written by the I/O device into memory, and then the data is either prefetched before access or demand fetched into the cache upon access by the processor.
With DDIO, however, DMA transactions for an I/O region go directly to the last level cache, and they will be in dirty mode and will get written back to memory only upon eviction~\cite{DavidKanter12,Farshin19ddio}.

While DCA and DDIO have been shown to improve packet processing speeds by reducing the cache miss rates in many scenarios~\cite{dca,ddio}, if the device has large descriptor rings, they could potentially degrade performance by evicting useful data out of the LLC~\cite{tang10dma}.
In addition, as we show in this paper, these technologies potentially open up new vulnerabilities since the packets are brought directly into the LLC,
which is shared by all cores in the processor. 
%
%
\vspace{-1ex}\subsection{Network-Based Covert-Channels}
The literature abounds with network-based covert channels that leverage network protocols as carriers by encoding the data into a protocol feature~\cite{zander2007survey,abad2001ip,girling1987covert,servetto2001communication,lucena2005covert,hintz2002covert}. 
For example, covert channels can be constructed by encoding data in unused or reserved bits of the frame or packet headers~\cite{lucena2005covert,handel1996hiding, hintz2002covert, kundur2003practical}, such as the TCP Urgent Pointer which is used to indicate high priority data~\cite{hintz2002covert}.
In the TCP protocol, Initial Sequence Number (ISN) is the first sequence number and is selected arbitrarily by the client.
Rowland~\cite{rowland1997covert} proposed shifting each covert byte by 26 bits to the left and directly using it as the TCP ISN.
Abad~\cite{abad2001ip} shows that the IP header checksum field can also be exploited for covert communication, and further proposes encoding the secret information into the checksum field and adding the content of an IP header extension to compensate the checksum modification, chosen such that the modified checksum will be correct. 
Other header fields such as address fields~\cite{girling1987covert} and packet length ~\cite{lucena2005covert} are also exploited to build covert channels. 
In addition to the header field, packet rate and timing~\cite{girling1987covert,kundur2003practical},  packet loss rate~\cite{servetto2001communication}, and packet sorting~\cite{cabuk2004ip} are also used to build covert channels. 

Many of these covert channels are based on non-standard or abnormal behaviour of the protocol and can be detected and prevented by simple anomaly detection methods~\cite{zander2007survey}.
%
%
%
In addition, all of these network-based covert channels require the receiver to have access to the network and be able to receive packets, while the receiver in our packet chasing attack does not need any access or permission to the network. 
\subsection{Cache Attacks} 
Cache-Based side-channel attacks are the most common class of architectural timing channel attacks, that leverage the cache as their sole medium of covert communication~\cite{bernstein2005cache,osvik2006,Wu12,Liu15}.  These attacks have the potential to reveal sensitive information such as cryptographic keys~\cite{Gruss2016,yarom2014flush+,craig,Liu15,CacheZoom,cachequote}, keystrokes~\cite{Gruss15,Wang2019UnveilingYK}, and web browsing data~\cite{Oren15,Shusterman19usenix}, by exploiting timing variations that arise as a result of a victim process and a spy process contending for a shared cache. 
For example, in the \textsc{Prime+Probe}~\cite{Liu15,Oren15} attack, the spy process infers the secret by learning the temporal secret-dependent cache access patterns of a victim, by contending for the same cache sets as the victim and measuring the timing variations that arise due to such contention. 
In the \textsc{Prime} step, the attacker fills one or more cache sets with its own cache blocks, simply by accessing its data. 
Then, in the \textsc{Idle} step, the attacker waits for a time interval and lets the victim execute and use the cache, possibly
evicting the attacker's blocks. 
Finally, in the \textsc{Probe} step, the attacker measures the time it takes to load each set of cache blocks.  If it
is noticeably slow, she can infer that the victim has accessed a block in that set, replacing the attacker's block. 
%
%

To perform these attacks in a fine time granularity, the attacker has to target specific sets in the last level cache. As such, she has to know how the addresses map into the sets in the LLC. 
\begin{figure}
    \centering
    \includegraphics[width=.48\textwidth]{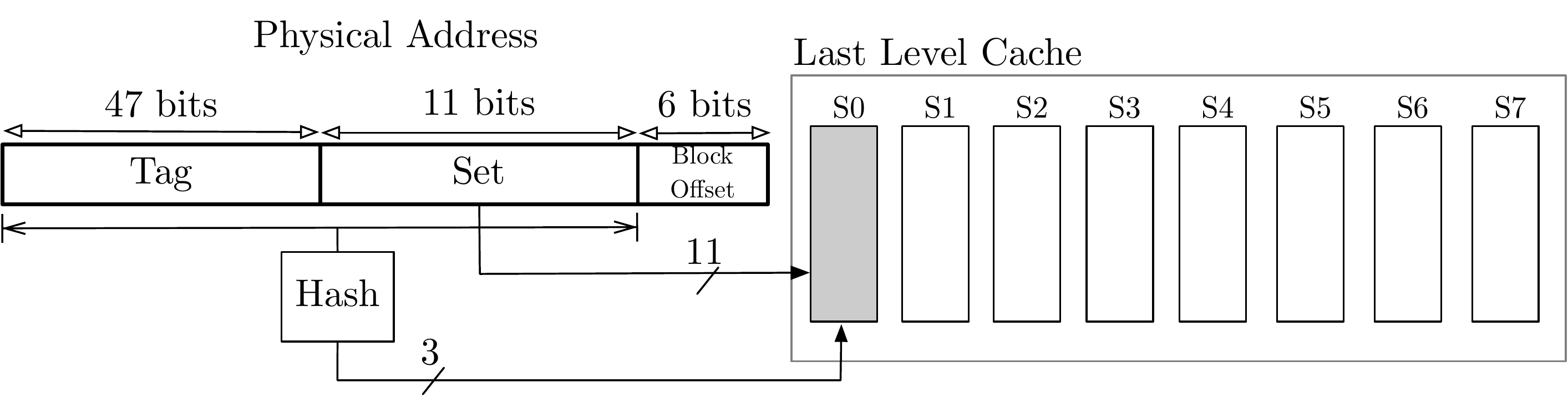}
    \caption{Intel's complex indexing of modern last level cache.}
    \label{fig:slice}
\end{figure}
However, starting with the Sandy Bridge microarchitecture~\cite{sandybridge}, Intel has employed a new LLC design, in which the LLC is split into multiple slices, one for each core (See figure~\ref{fig:slice}), with an unpublished hash function mapping physical addresses to slices, supposedly distributing the physical addresses uniformly among the cores. This hash function has since been successfully reverse-engineered for many different processors,
including Intel’s Sandy Bridge~\cite{kayaalp2016high,maurice2015reverse,yarom2015mapping}, Ivy Bridge~\cite{maurice2015reverse,inci2016cache}, and Haswell~\cite{maurice2015reverse,irazoqui2015systematic} architectures. 

\ignore{{\color{red} If we can do it without being obvious, it might be 
nice to cite prime+abort.  Maybe a sentence to mention other 
cache side channel attacks such as flush+reload, etc.? Don't
force it, though. --How about the following? }
}

In addition to \textsc{Prime+Probe},  multiple other variants of cache attacks are also proposed~\cite{yarom2014flush+, craig, Gruss2016}. \textsc{Flush+Reload}~\cite{yarom2014flush+} uses Intel’s CLFLUSH instruction to flush a target address out of the cache,  and then, at the measurement phase, the attacker “reloads” the target address and measure its access time. 
However, it relies on shared memory between the spy and the victim, and requires access to precise timers. 
\textsc{Prime+Abort}~\cite{craig} exploits Intel's transactional memory extension (TSX) hardware to mount a timer-free last level cache attack.  

Several defenses have been proposed in the literature to mitigate cache timing channels~\cite{Gururaj,oberg2013practical,mao2017quantitative,Li14Sapper,Kayaalp17dac,Wang16SecDCP,Wang2007,csf,Wang17CacheD,zhang2016cloudradar,kiriansky2018dawg,Varys18,ghostrider,raccoon,csd,csd_top}.  
These mitigation strategies include identifying the leakage in software~\cite{Wang17CacheD}, observing anamalous cache behavior~\cite{zhang2016cloudradar,taram2019fast}, closing channels at hardware design time~\cite{Wang16SecDCP,Kayaalp17dac,oberg2013practical,mao2017quantitative,Li14Sapper}, dynamic cache partitioning~\cite{Wang2007,kiriansky2018dawg,Wang16SecDCP}, 
strictly reserving physical cores to security-sensitive threads~\cite{Varys18}, randomization~\cite{Wang2007}, memory trace obliviousness~\cite{ghostrider,raccoon}, and cache state obfuscation using decoy load micro-ops~\cite{csd,csd_top}. 

\subsection{Security of I/O Devices and Drivers}
A number of security attacks have been published that target device drivers~\cite{zhou2014peril,ThunderclapNDSS2019,gorobets2015attacking}. Thunderclap~\cite{ThunderclapNDSS2019} describes an attack that subverts the Input-Output Memory Management Unit (IOMMU) protections to expose the shared memory available to DMA-enabled I/O peripherals.
Zhu, et al.~\cite{Zhu17} demonstrate another attack that bypasses IOMMU and compromises the GPU driver to exploit GPU microcode to gain full access to CPU physical memory.
 To address these vulnerabilities, researchers focus on isolating device drivers, and to make operating systems secure when a device driver is buggy or has code which is intentionally malicious~\cite{Boyd10Malicious,Tan07ikernel}. 
 Tiwari, et al.~\cite{Tiwari11Microkernel} propose a full system which includes an I/O subsystem and a micro-kernel that enable isolation and secure communication by monitoring and controlling the information flow of the system.

{ NetCat~\cite{netcat} is a concurrent work to our Packet Chasing attack.  It describes an attack that exploits a similar underlying vulnerability. However, this work differs in many important ways.
First, NetCat only detects the arrival time of packets, whereas Packet Chasing has the ability to detect
both arrival time and size of each packet -- the latter is more reliable and less 
noisy.  This gives Packet Chasing-based attacks the opportunity to mount more powerful attacks such as the web fingerprinting attack that we describe in this paper (Section~\ref{sec:side}). 
Second, unlike Packet Chasing, NetCat requires DDIO and RDMA technologies to be present, limiting its generality.
Therefore, to mitigate NetCat, it is sufficient to disable DDIO or RDMA.  However, as we show in this paper, the Packet Chasing attack is practical even in the absence of those technologies. Therefore, we also present a more sophisticated yet high-performance defense that mitigates the attacks. } 

\section{Packet Chasing: Setting up the Attack}\label{sec:setup}

We perform our analysis and attack on Intel's Gigabit Ethernet (IGB) driver version 5.3.5.22~\cite{igb} loaded into Linux Kernel version 4.4.0-142. 
We run the attack on a Dell PowerEdge T620~\cite{dell} server which uses Intel I350 network adapter~\cite{i350} and is operated by two Intel Xeon CPU E5-2660 processors. Each processor has a 20 MB last level cache with 16384 sets. 
To perform \textsc{Prime+Probe} on the last level cache, we use the Mastik Micro-Architectural Side-Channel Toolkit Version 0.02~\cite{mastik}

Our attack consists of two phases. One is an offline phase where the attacker recovers the sequence of the buffers and an online phase where the attacker uses that information to monitor the incoming packets. 

\subsection{ Deconstruction of the NIC Driver}\label{sec:deconstruction}
While the code samples of this subsection are specific to Intel's Gigabit Ethernet (IGB) driver, we note that the insights are generalizable. 
%
The original Ethernet IEEE 802.3 standard defines the minimum Ethernet frame size as 64 bytes and the maximum as 1518 bytes, with the maximum being later increased to 1522 bytes to allow for VLAN tagging. 
Since the driver and the NIC don't know the size of incoming packets beforehand, the NIC has to allocate a buffer that can accommodate any size. 
The IGB driver allocates a 2048 byte buffer for each frame and packs up to two buffers into one 4096 byte page which will be synchronized with the network adapter. 
For compatibility, it is recommended~\cite{dma} that when the device drivers map a memory region for DMA, they only map memory regions that begin and end on page boundaries, which are guaranteed also to be cache line boundaries.
%
Further, the \textit{rx} ring buffer is used to temporarily hold packets while the host is processing them. While employing more buffers in the ring could reduce the packet drop rate, it could also increase the host memory usage and the cache footprint. Therefore, although the maximum size supported by Intel's I350 adapter is 4096 buffers, the default value in the IGB driver is set to 256.  

The linux kernel, in the DMA API, provides two different types of DMA memory allocation for device drivers. 
Coherent (or consistent) memory and streaming DMA mappings. Coherent memory is a type of DMA memory mapping for which a write by either the device or the processor can be visible and read by the processor or device without the need to explicitly synchronize and having to worry about caching effects. 
However, the processor has to flush the write buffers before notifying devices to read that memory~\cite{dma}.
Therefore, consistent memory can be expensive on some platforms as it invariably entails a wait due to write barriers and flushing of buffers~\cite{dma}. 
While the buffers themselves are mapped using streaming DMA mapping, the ring descriptors are mapped using coherent memory. Thus, the device and the driver have the same view of the ring descriptors. %
Also, this makes the writes to the \textit{rx} descriptor ring expensive. Therefore, in order to avoid changing the content of \textit{rx} descriptors, drivers after receiving packets usually reuse the buffers instead of allocating new buffers. 
So the drivers usually allocate the buffers once and reuse them throughout the life cycle of the driver.

\begin{figure}
\begin{minted}{c}
static bool igb_add_rx_frag(rx_buffer, skb){
    ...
    size = rx_buffer->size;
    page = rx_buffer->page; 
    if (likely(size <= IGB_RX_HDR_LEN)) {
        memcpy(__skb_put(skb, size), page, size);
        /* we can reuse buffer as-is,
        just make sure it is local */
        if (likely(page_to_nid(page) == numa_node_id()))
            return true;
        /* this is a remote page and cannot be reused*/
        put_page(page);
        return false;
    }
    /* only if packet is large */
    skb_add_rx_frag(skb, page);
    return igb_can_reuse_rx_page(rx_buffer, page);
}        
\end{minted}
\caption{The IGB driver function that adds the contents of an incoming buffer to a socket\_buffer which will be passed to the higher levels of networking stack. The function returns true if the buffer can be reused by the NIC.}
\label{fig:igb_add}
\end{figure}

\begin{figure}
\begin{minted}{c}
bool igb_can_reuse_rx_page(rx_buffer, page){  
  /* avoid re-using remote pages */
  if (unlikely(page_to_nid(page) != numa_node_id()))
      return false;                                                                                 
  /* if we are only owner of page we can reuse it */
  if (unlikely(page_count(page) != 1))
      return false;                                                             
  /* flip page offset to other buffer */
  rx_buffer->page_offset ^= IGB_RX_BUFSZ;
  /* bump page refcount before it's given to stack */
  get_page(page);                                             
  return true;
}
\end{minted}
\caption{The IGB driver function that checks if the driver can reuse a page and put it back into the rx ring buffer.}
\label{fig:igb_reuse}
\end{figure}

Figure~\ref{fig:igb_add} shows the part of the IGB driver code that is called upon receiving packets and whose job is to add the contents of the \textit{rx} buffer to the socket buffer which will be passed to the IP layer.  If the size of the packet is less than a predefined threshold (256 by default), then the driver copies the contents of the buffer and then tries to recycle the same buffer for future packets.
If the buffer is allocated on a remote NUMA node, then the access time to that buffer is much more than if the buffer was allocated in a local NUMA node. Therefore, to improve performance, the driver deallocates the remote buffer and re-allocates a new buffer for that \textit{rx} ring descriptor. 
If the packet size is larger than 256, then instead of the direct copy, the IGB driver attaches the page as a fragment to the socket buffer. It then calls the \textit{igb\_can\_reuse\_page} function shown in Figure~\ref{fig:igb_reuse}.  This function checks for two conditions that are unlikely to be met.  
The first condition is that the buffer is allocated on a remote NUMA node.  The second condition is that the kernel is still preparing the packet in the other half and that the driver is not the sole owner of the page. 
If neither condition is met, the driver flips the \textit{page\_offset} field, so that the device only uses the second half of the page.

To summarize, in the common scenarios, the driver uses a small number of ring buffers (256) on 256 distinct pages,
each of them half-page aligned and it continually reuses these buffers typically
until the next system reboot or networking restart. 
In addition, to maintain high (and consistent) packet processing speeds, the order of the ring descriptors does not change throughout the execution of the driver code.  Therefore, as long as the driver reuses the buffers for descriptors, the order of the buffers remains constant.

\subsection{Recovering the Cache Footprint of the Ring Buffer}
The ultimate goal of the Packet Chasing attacker is to gain size and temporal information about incoming packets by spying on the last level cache.  
To this end, we mount a \textsc{Prime+Probe} attack on the last-level cache. However, blindly probing all cache sets doesn't give us much information. 
This is because the probe time is limited by the time it takes to access the entire cache, which in this case is about 12 million CPU cycles, too long to gain any useful information about incoming packets.
Long probe time also makes the attack more susceptible to background noise, as the probability of observing irrelevant activity on the cache line increases. 

However, from the previous subsection, we know that the buffers that store packets in kernel memory are page-aligned. 
That means we only need to probe the sets that the page-aligned addresses are mapped to. 
Having 4KB page size implies that the lowest 12 bits of the starting addresses are zero. So the lowest 6 bits of the set indices are zero (also see Figure~\ref{fig:slice}). That limits us to 32 sets in each slice for a total of 256 possible sets. 
Using the Mastik toolkit, we find these sets and construct eviction sets for them, which are essentially a stream of addresses
guaranteed to replace all other data from all the cache blocks in
a set.
%
With these, we have the ability to monitor all 256 cache sets that are potential candidates for buffer locations.

\begin{figure}
    \centering
    \includegraphics[width=.49\textwidth]{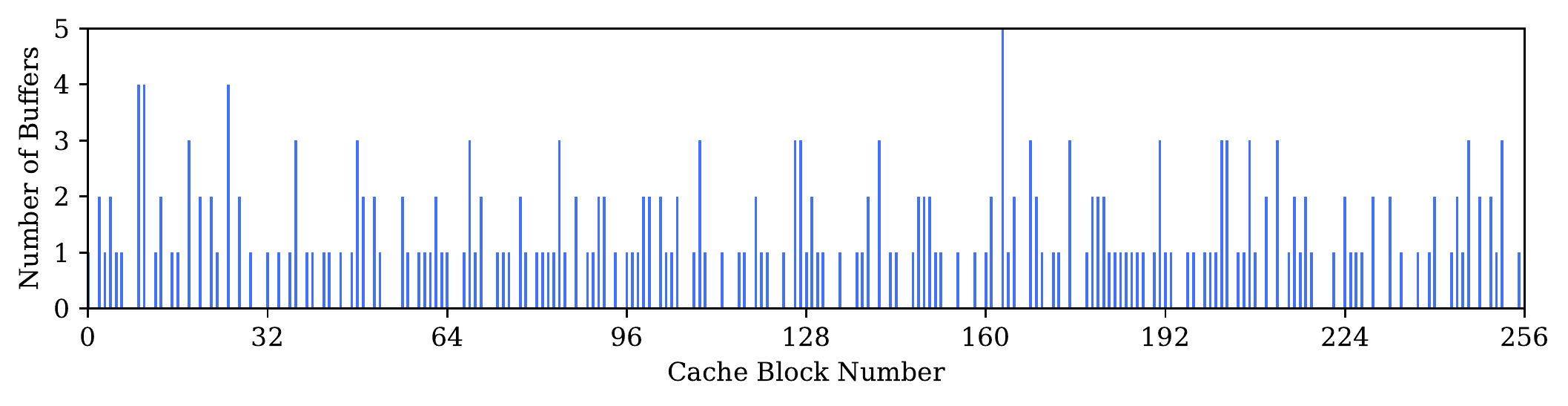}
    \caption{An example of how the NIC ring buffers are mapped to  to the page-aligned cache sets.}
    \label{fig:doubling-example}
    \vspace{-2ex}
\end{figure}

While all the NIC \textit{rx} buffers map to one of the page-aligned cache sets that we obtain, the distribution of this mapping is not uniform, which means that some of the \textit{rx} buffers are mapped to the same cache set. 
To show an example of such conflict in the cache sets, we instrument the driver code to print the physical addresses of the ring buffers, which we then map to cache set indices. 
Figure~\ref{fig:doubling-example} shows this non-uniform mapping for just one instance of the buffer allocation in the NIC. 
Horizontal axes shows one of the page-aligned cache sets and on the Y axis, we show the number of NIC buffers that map to each page-aligned cache set. 
In this example, we see that 5 NIC buffers are mapped to cache set number 165 while none of the NIC buffers are mapped to cache block 65.  

\begin{figure}
    \centering
    \includegraphics[width=.45\textwidth]{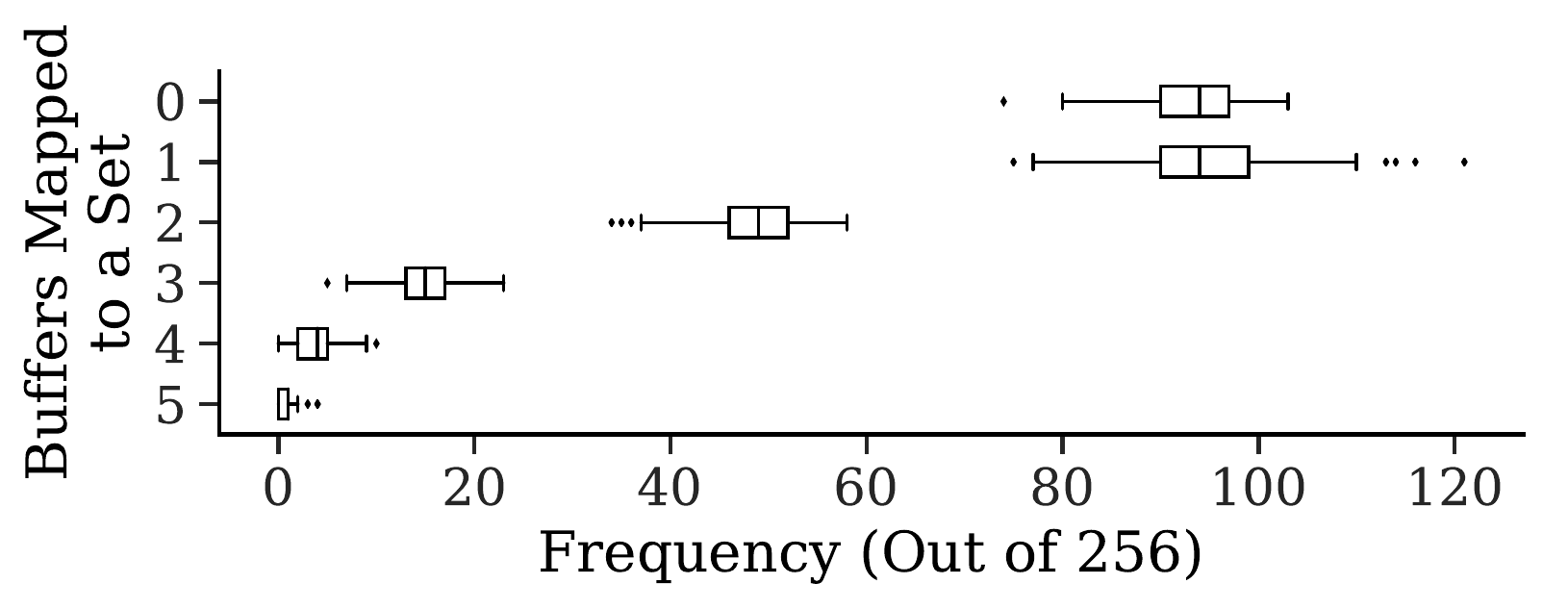}
    \caption{Frequency of the ring buffers that map to same sets, measured for 1000 instances. Zero represents the number of sets that are page aligned but none of the ring buffers is mapped to those.}
    \label{fig:doubling-histogram}
\end{figure}

Figure~\ref{fig:doubling-histogram} further analyzes this mapping which shows the result of performing the same experiment across multiple instances of driver initialization. 
%
%
For around 35\% of the page-aligned sets, there is no co-mapped NIC buffer, while there are only 5 out of 1000 instances in which we see more than 4 buffers mapped to the same page-aligned cache set. 

\begin{figure}
    \centering
    \includegraphics[width=\linewidth]{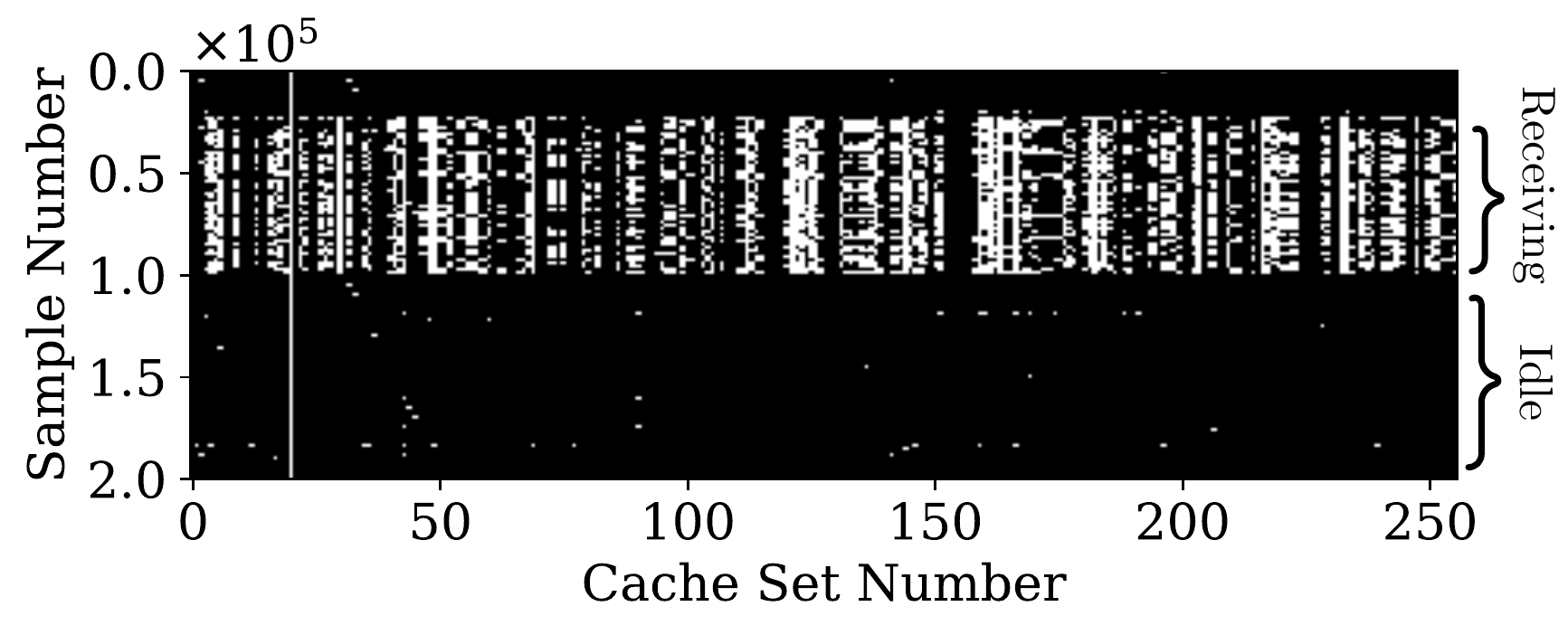}
    \caption{Monitoring all the page-aligned sets while receiving packets. A white dot shows at least one miss (activity) on a cache set in a sample interval}
    \label{fig:idle}
\end{figure}

By narrowing down the number of monitored cache sets to only the 256 possible buffer starting
locations, we are able to see a clear footprint in the cache when the NIC device is receiving packets, as shown in Figure~\ref{fig:idle}.
In this experiment, we rely on a remote sender who is on the same network with the spy and constantly sends broadcast Ethernet frames to the network. To this end, we use Linux raw socket~\cite{rawsocket} which generates broadcast Ethernet frames with arbitrary sizes. These frames get discarded in the driver since the protocol field is unknown. Thus, the effect that we see is only caused by the driver/adaptor accessing the buffers, without any activity of the kernel networking stack. At around sample 25k, the sender starts sending packets and it continues to do so until sample 100k.  
In some cache sets, e.g., cache set number 53, we don't see any activity and that is because none of the NIC buffers are mapped to those sets.\ignore{ On the other hand, on some cache sets, e.g., cache set number 21, we see activity regardless of whether the packets are arriving at the NIC, due to the always-miss scenario described in Section~\ref{sec:setup}.  }

 The packet chasing attacker, with the ability to distinguish between an idle system vs. when there are incoming packets, establishes a leaking channel that can be exploited to covertly communicate secret data over the network. 
 We can further increase the bandwidth of this channel by differentiating the receiving streams based on frame sizes. 
 Since the incoming packets are stored in contiguous \textit{rx} buffers, using the same way that we construct the eviction sets for the page-aligned cache sets, we construct eviction sets for the second cache blocks in the page. 
 All the second cache blocks in the pages are mapped to one of these 256 cache sets. 
 Similarly we find the sets for the third and fourth cache blocks of the pages. This now allows us to recognize not just the presence of
 a packet, but also the size of the packet.
 
 \begin{figure}
    \centering
    \includegraphics[width=0.49\textwidth]{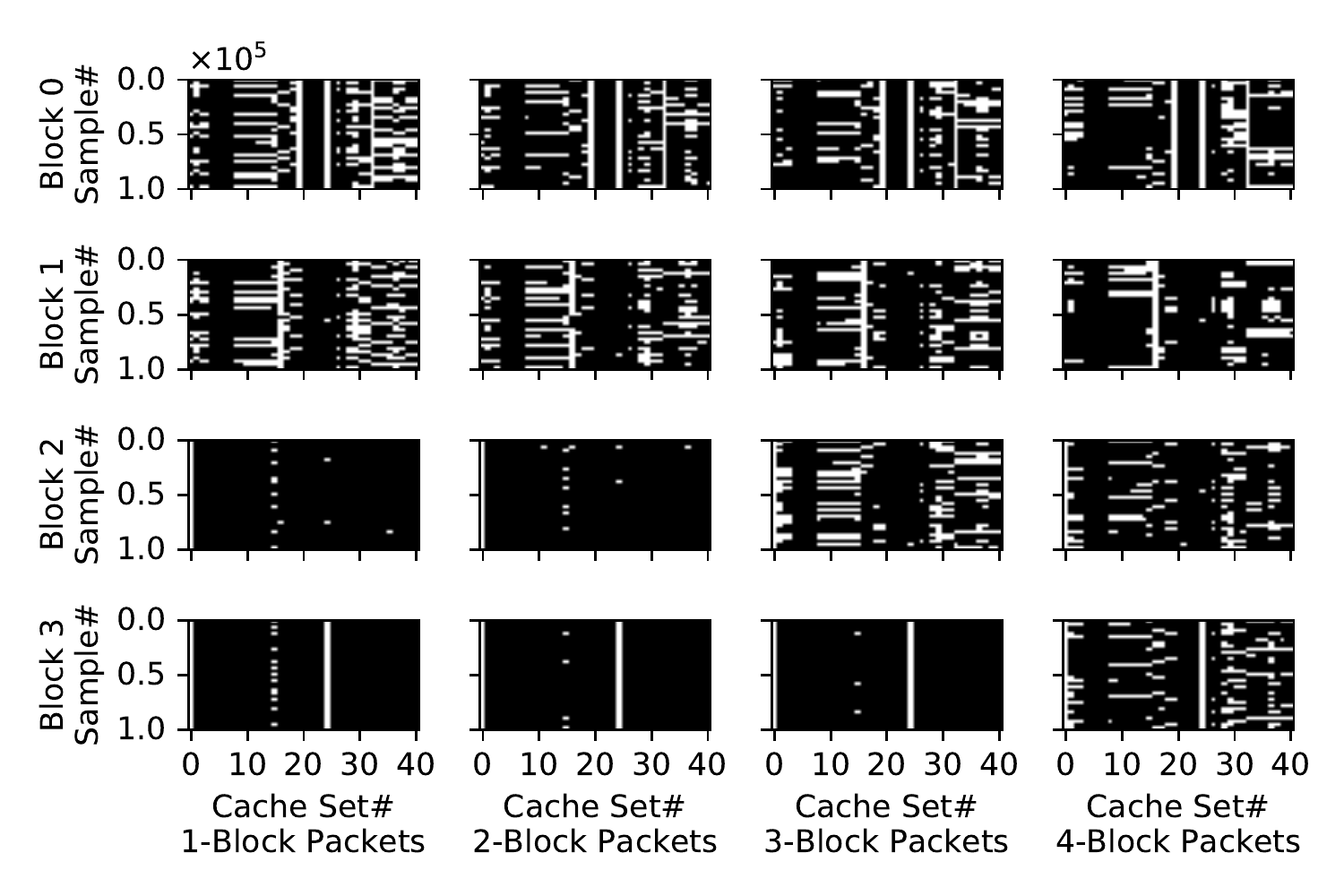}
    \caption{Cache footprint of packets with different sizes while probing the addresses that map to the location of the first three blocks in the packet buffer page. A white dot indicates at least a miss in a set.}
    \label{fig:size}
    \vspace{-2.6ex}
\end{figure}

 Figure~\ref{fig:size} shows the result of a simple experiment where
 we send packets of different sizes and test our ability to detect
 packet size.  On the columns, we have four different runs with
 constant packet sizes being sent, from one cache block (64 bytes) to four cache blocks (256 bytes). 
On the rows we show detection on four different cache eviction sets, block 0 to block 3 which are targeting the first to fourth blocks in the page-aligned buffers. 
As expected, we see clear activity on the diagonal and above, and
no activity below the diagonal.
 The only exception is 1-block packets which exhibit activity on block 1 as well as on block 0.
 This is because there is a performance optimization in the driver code that prefetches the second block of the packet regardless of the packet sizes. 
 The reason for this optimization is that most Ethernet packets have at least two blocks, and 64-byte packets (0-Block Packet) are only common in control packets that don't have payloads such as TCP acknowledge packets. 

The attack distinguishes a stream of packets with different sizes from each other, and that could be used to construct a remote covert channel (more details in Sec~\ref{sec:covert}) with 1950 bytes-per-second bandwidth by only detecting a stream of small packets vs. a stream of large packets (essentially, a binary signal).
However, we can turn this to a more powerful channel if we differentiate sizes with finer granularity, essentially sending
multiple bits of information per packet.
The following subsection describes the method that we use to further narrow down the monitored sets while we perform \textsc{Prime+Probe}. 

\subsection{Chasing Packets over the Cache}\label{sec:sequence}

The attacker has to probe all 256 page-aligned sets at once to detect incoming packets only because she doesn't know which buffers get filled first, and then probe more sets to detect packet size.
However, if we know the order in which the buffers get filled in the driver, then we can actually chase the packets over the cache by only probing the cache sets corresponding to the next expected buffer, building a powerful high-resolution attack. 
We show that it is possible to almost fully recover the sequence of the buffers, in a
one-time statistical analysis phase.  
Since the buffers are always recycled and then returned to the ring, the order of the buffers in the ring is maintained during the lifetime of the driver. 

Algorithm~\ref{alg:sequencer} describes the \textsc{sequencer} procedure that we use to recover the sequence. 
It consists of three steps. First, in the \textsc{Get\_Clean\_Samples} step, we gather cache probe samples for \emph{Nsets} cache sets. 
To this end, we start with constructing the eviction sets for the page-aligned NIC buffers.
However, sometimes we have always-miss scenarios on some sets, which is easily observed a priori.
For those sets, we simply use the second cache block of a page-aligned buffer instead of the first one. 

\begin{figure}
    \centering
    \includegraphics[scale=.18]{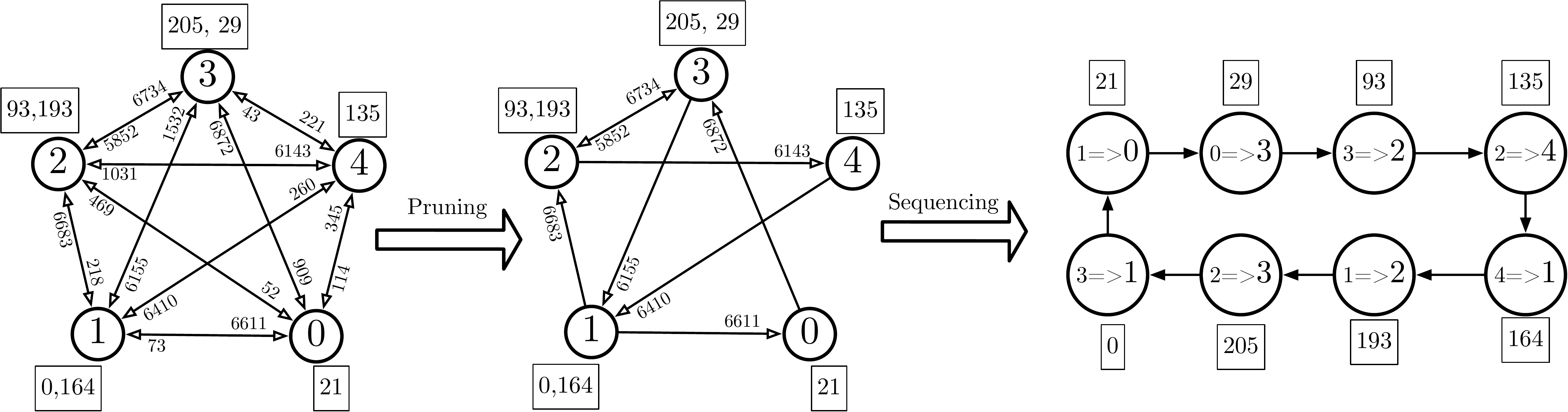}
    \caption{Pruning and sequencing of the set graph to get the order of ring buffers. Each node represents a set in the attacker address space. Numbers in squares are the sequence number of the associated ring buffers that map to same set.}
    \label{fig:graph}
\end{figure}

\setlength{\textfloatsep}{0pt}%
 \begin{algorithm}[t]
\footnotesize
\caption{Ring Buffer Sequence Recovery }\label{alg:sequencer}
\newcommand\Algphase[1]{%
\vspace*{-.7\baselineskip}\Statex\hspace*{\dimexpr-\algorithmicindent-2pt\relax}\rule{\linewidth+\algorithmicindent}{0.4pt}%
\vspace*{-.2\baselineskip}
\Statex\hspace*{-\algorithmicindent}\textbf{#1}%
\vspace*{-.7\baselineskip}\Statex\hspace*{\dimexpr-\algorithmicindent-2pt\relax}\rule{\linewidth+\algorithmicindent}{0.4pt}%
}
\begin{algorithmic}[1]
\Procedure{Sequencer}{}
    \State{samples $\gets$ \textsc{get\_clean\_samples}(Nsets, Nsamples)}
    \State{graph   $\gets$ \textsc{build\_graph(samples)}}
    \State{sequence  $\gets$ \textsc{make\_sequence(graph)}}
    \State{\textbf{return} sequence}
\EndProcedure
\Procedure{get\_clean\_samples}{Nsets, Nsamples}
\State monitor\_list $\leftarrow [0..Nsets]$
\State samples $\leftarrow$ repeated\_probe(Nsamples, monitor\_list)
\ForAll{ x $\in$ monitor\_list} \If{activity(samples[x]) $>$ activity\_cutoff}  \State{replace x in monitor\_list 
with the 2\textsuperscript{nd} block of the page }
\State{\textbf{goto: 3}}
\EndIf
\EndFor
\State{\textbf{return} samples}
\EndProcedure
\Procedure{build\_graph}{samples}
\State{curr$\gets0$, prev$\gets0$}
\For{i $\in \{0,...,SAMPLES\}$} 
\ForAll{ cand $\in$ monitor\_list} 
\If{samples[i][cand] $<$ miss\_threshold} \Comment{no activity}
\State{\textbf{continue}}
\EndIf
\If{curr$\neq$ prev} \Comment{no self-loop}
\State{graph[prev][curr][cand]$\gets$ graph[prev][curr][cand]+1}
\State{(prev, curr)$\gets$(curr, cand)}
\EndIf
\EndFor
\EndFor
\State{\textbf{return} graph}
\EndProcedure
\Procedure{make\_sequence}{graph}
\State root $\gets$ get\_root(graph)
\State sequence $\gets []$, (prev,curr) $\gets$ root
\Repeat
\State sequence.push(curr)
\State{(next,weight)$\gets$ get\_max\_weight(graph[prev][curr])}
\If{weight $<$ weight\_cutoff}
\State{\textbf{break}}
\EndIf
\State{graph[prev][curr][next] $\gets$ 0}\Comment{mark as visited}
\State{(prev, curr) $\gets$ (curr, next)}
\Until{(prev, curr) $\neq$ root}
\State{\textbf{return} sequence}
\EndProcedure
\end{algorithmic}

\end{algorithm}

After that, we start building a complete weighted graph with the nodes being the monitored cache sets and the weights on the edge that connect node \textit{x} to node \textit{y} are the number of times that we observe an activity on set \textit{y} which was immediately followed by an activity on set \textit{x}, as illustrated in the leftmost graph in Figure~\ref{fig:graph}. 
To deal with the problem that multiple buffers can map to the same cache set, when we build the graph, we maintain one node history for each edge.
This allows the algorithm to distinguish between the activity on two or more different buffers that map to the same cache set by their successor cache sets.
So, for example in Figure~\ref{fig:graph}, two different buffers are mapped to cache set number 2.  These buffers occupy location numbers 93 and 193 in the ring buffer. 
Therefore, in the final sequence, we have two different instances of cache set number 2, one that is followed by cache set 3 and the other that is followed by cache set 1.

The final step, \textsc{Make\_Sequence}, is to traverse the graph we build in the previous steps, starting from a random node, and continuing to move forward until we reach the same node.
Note that since the final sequence is a \textit{ring} in which the \textit{in-degree} and the \textit{out-degree} of each node is exactly one, the choice of the starting node doesn't change the outcome. 

While this procedure can recover the sequence of the buffers that are mapped into \textit{Nset}, it can only do so if we monitor a limited portion of the page-aligned cache sets (we were successful up to 64 cache sets).  This is because the probe time gets longer than what is required to detect the order of the incoming packets, if we include more sets in our monitor list. 
So we first find the sequence for 32 cache sets, then we repeat the \textsc{Sequencer} procedure with the first 31 nodes (node 0 to node 30) plus a candidate node (e.g, 32) and we try to find the location of the candidate in the sequence. 
Then, we repeat the same procedure, moving through the node sequence, until we find a place in the sequence for all cache sets. 

Sometimes two consecutive buffers are mapped into one set. For example, consider the case that buffers number 93 and 98 are mapped into the set 2 in Figure~\ref{fig:graph}. With our approach, we don't capture these cases in the first round, but starting from the beginning, when we do encounter a
buffer that is between the two, we can split the two in our graph.  
If they are truly consecutive
in the final ring (unlikely), the buffers are essentially merged, but this has no impact on our
mechanism to create a covert channel, and will have minimal effect on the overall fingerprint we
observe in the web fingerprinting leakage attack.

We measure Levenshtein distance~\cite{jm} to quantify the distance between the sequence that we obtain and the ground truth actual sequence that we get from driver instrumentation. The Levenshtein distance between two sequences is the minimum number of single-character edits (i.e., insertions, deletions or substitutions) needed to change one sequence into another.  We see the results of this experiment in Table~\ref{table:seq}.   
\ignore
Fine-tuning the probe rate is a rather challenging task as it needs to be long enough that the activity of each incoming packet touches only one sample, and needs to be small enough to not lose the temporal relation between the incoming consecutive packets. 
Otherwise, we see a drop in accuracy of the obtained sequence. However, in our covert-channel construction, in many of our attack scenarios, we only need to
find buffers that are sufficiently far apart in the ring, so small errors in the sequence are tolerable. 

During the profiling period we rely on a remote sender whose only job is to constantly send packets. 
However, the spy can recover the sequence even without the help of the external sender, as long as the system is receiving packets.
In fact, noise (extra packets not sent by co-operating sender) in
this step only helps the spy.

\setlength{\tabcolsep}{1em}
\begin{table}
\footnotesize
\centering
\vspace{-1ex}
\caption{Summary of experiments for sequence recovery}\label{table:seq}
\begin{tabular}{|lll|}
  \hline
   \multicolumn{3}{|c|}{\textbf{Results}} \\ \hline
    Measure  & Value & CI  \\ \hline
Levenshtein Distance & 25.2 & {[}22, 35{]}\\
 Error Rate (\%) & 9.8 & {[}8.5, 13.6{]} \\
Longest Mismatch & 5.2 & {[}3, 9{]}  \\ 
Time (Minutes)        &159   & {[}153, 167{]}\\
\hline 
\multicolumn{3}{|c|}{\textbf{Parameters}}\\ \hline
 Parameter & \multicolumn{2}{l|}{{Value}}  \\ \hline
 Number of Samples   &\multicolumn{2}{l|}{100,000} \\
 Number of Monitored Set   &\multicolumn{2}{l|}{32} \\
Packet Rate (packet/s)  &\multicolumn{2}{l|}{ 0.2M} \\
Probe Rate (probe/s) &\multicolumn{2}{l|}{8000} \\ \hline
\end{tabular}
\end{table}

\section{Packet Chasing: Receiving Packets without Network Access}\label{sec:covert}
In this section, we show the effectiveness of the Packet Chasing attack by constructing a covert channel over the network. 
We assume a simple threat model where a remote trojan attempts to send covert messages through the network, to a spy process located in the same physical network. 
 The spy process can be inside a container and does not have root privileges, neither in the container nor in the host OS, and is also not permitted to use the networking stack. 
The trojan process has the ability to send packets to the physical network, but there is no authorized method to communicate with the spy. 
\paragraph{Channel Capacity} 
To build a framework for quantitatively comparing  different encoding and synchronization schemes, we follow the methodology described by Liu, et al.~\cite{Liu15} to measure our channel bandwidth and error rate, while transferring a long pseudo-random bit sequence with a period of $2^{15} -1$. 
The pseudo-random bit sequence is generated using a 15-bit wide linear feedback shift register (LFSR) that covers all the $2^{15}$ sequences, except the case that all bits are zeros. 
This allows us to spot various errors that might happen during the transmission including bit loss, multiple insertion of bits, or bit swaps~\cite{Liu15}.
\paragraph{Data Encoding and Synchronization}
The spy first chooses $x$,  one of the page-aligned cache sets that only one of the ring buffers is mapped to.
Finding such a page-aligned cache set is not challenging using the approaches described in Section~\ref{sec:setup}.
Then the spy process finds the cache sets to which the addresses $x+64$, $x+2*64$, and $x+3*64$ are mapped. In other words, it finds the cache sets for the second, third, and fourth cache blocks of the page-aligned buffer. 
As described in Section~\ref{sec:setup}, the spy process knows the set index bits for these sets, but the outcome of the hash function (slice bits) is not known. 
To find out the exact slice, the spy process executes a trial and error procedure in which it selects one of the eight slices based on the activity on the sets.
After this step, the initialization is done and the spy process constantly monitors the found cache lines. 

The spy process, at time frame $n$, sends 256 (the length of the ring buffer) packets of size $(S+2)*64$ to transmit the symbol S.
Since we operate on the network and the latency is fluctuating frequently, we cannot use return-to-zero self-clocking encoding~\cite{Liu15}, rather we choose to use a synchronized clock encoding scheme in which the first block of the buffer acts as a clock to synchronize the spy with the trojan. 
We measure the bandwidth and the error rate for two cases.  First, we encode one binary symbol in each packet, i.e., we send either 64-byte packets that encode "0", or we send 256-byte packets that encode "1". Second, we send a ternary symbol in each packet, i.e, we send 64-byte packets to encode "0", 192-byte packets to encode "1", and 256-byte packets to encode "2". 

\begin{figure}
    \centering
    \includegraphics[width=.43\textwidth]{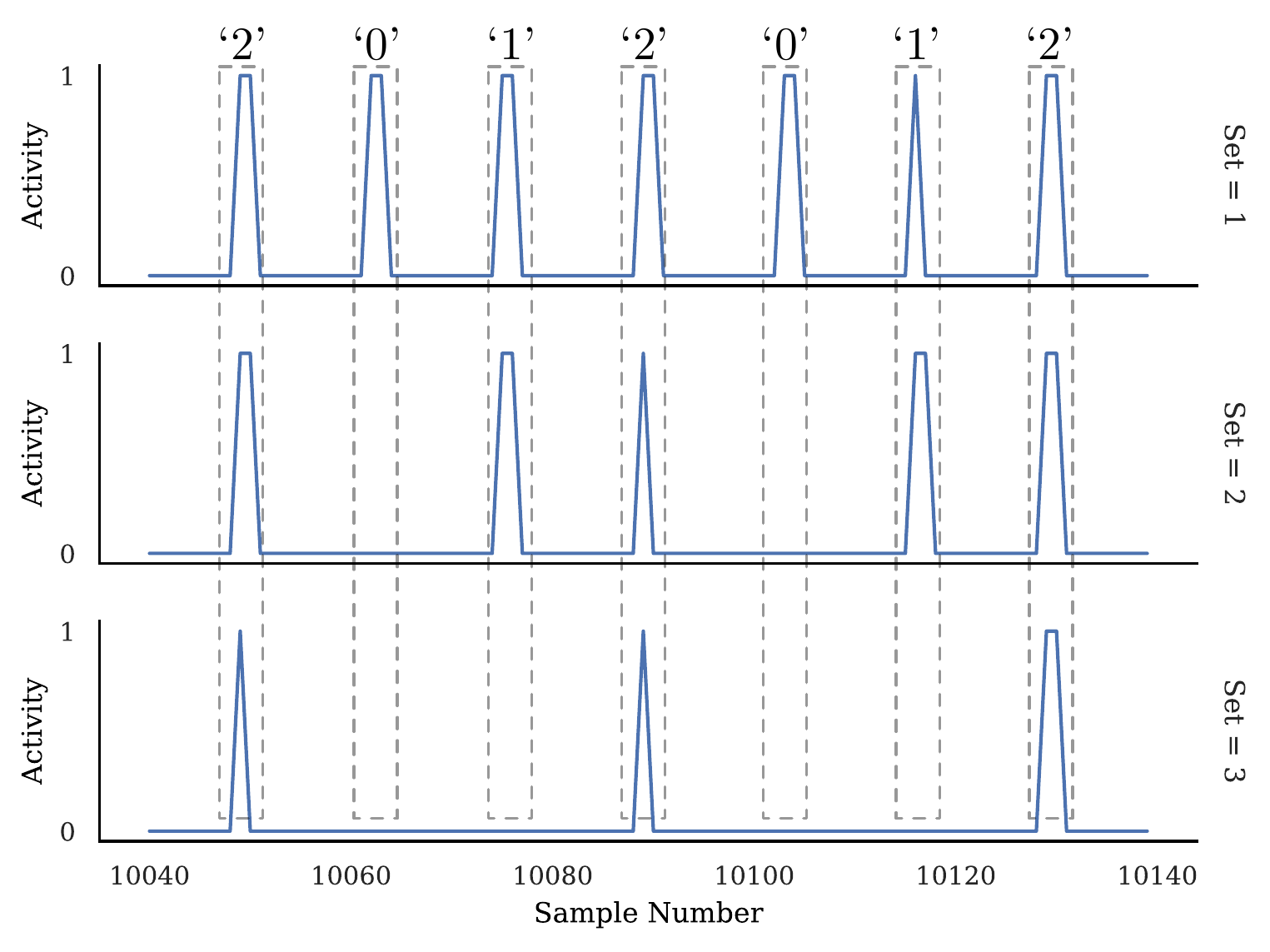}
    \vspace{-1ex}
    \caption{Spy process decodes the transmitted sequence based on the monitored activity on the probed sets. Set 1 acts as the clock and the activity is one for a set if we find at least one miss in the blocks in the eviction set of the probed set. }
    \label{fig:covert-sample}
\end{figure}

For example, Figure~\ref{fig:covert-sample} shows a part of a sequence that the spy receives in a real experiment.
In this experiment, the trojan transmits sequence "2012012012..." and the spy collects one sample from the three cache sets, every $200,000$ cycles.
When decoding, the spy uses a window of three samples to distinguish between different values. This is because sometimes we see the cache activity of one packet (one symbol) that spans across two cycles (the wide peaks in the figure).  The spy process should not decode these cases into two different symbols. In addition, having a window helps if the activity on the sets get skewed because of the delay of arriving packets.  
The first set of the buffer is used as a clock to synchronize the parties, and activity on the other two sets can show the transmitted values. Monitoring the activity of the two sets only gives us three different symbols because by sending a 3-block packet, we have a compulsory activity on set 2. 
{}

\begin{figure}[!t]
  	\centering
  \begin{subfigure}{0.235\textwidth}
  	\centering
  	\includegraphics[width=\textwidth]{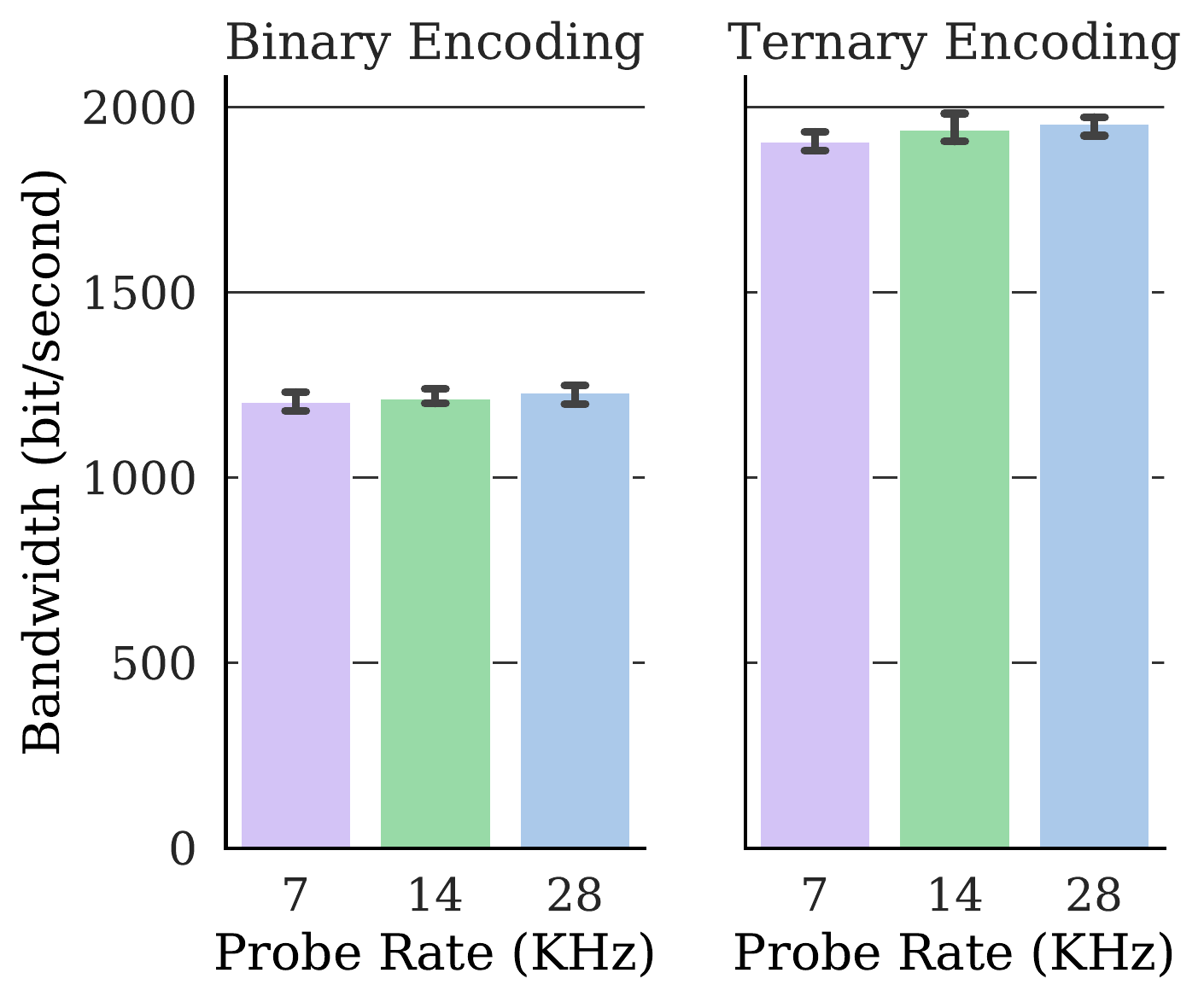} 
    \caption{Bandwidth }
  \end{subfigure}
  \begin{subfigure}{0.235\textwidth}
  	\centering
  	\includegraphics[width=\textwidth]{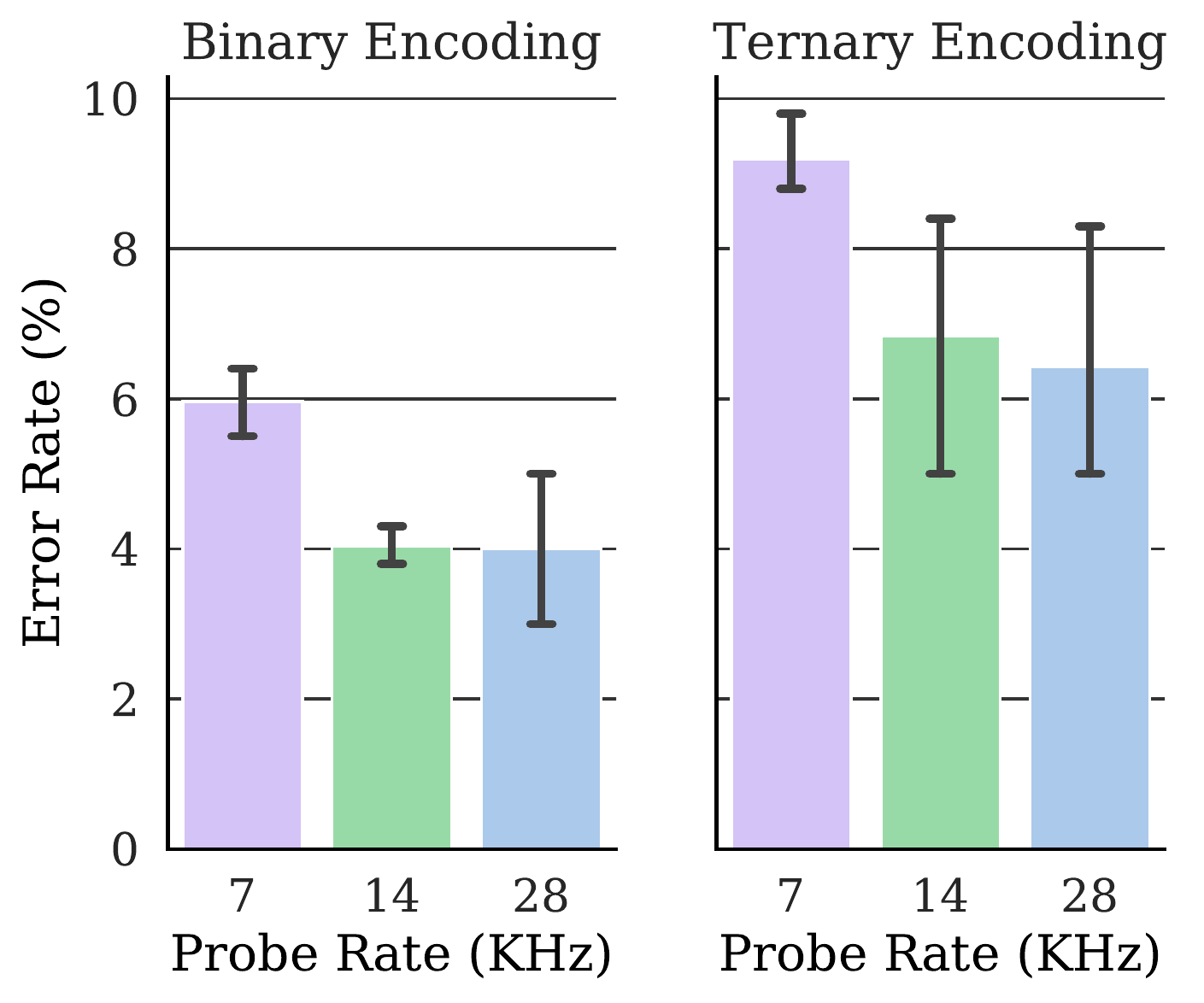} 
    \caption{Error Rate}
  \end{subfigure}
  \caption{Bandwidth and error rate of the remote covert channel for binary and ternary encoding and various cache probe rates. }
  \label{fig:covert_results}
\end{figure}

To estimate the error rate, we again use edit Levenshtein distance~\cite{jm} between the sent and received data for the pseudo random bit sequence.
Figure~\ref{fig:covert_results} shows the bandwidth and the error rate of our coding schemes, as well as the effect of varying the probe rate, i.e., the time we wait between consecutive probes. The bandwidth of the channel is almost constant with different probe rates.
This is because the limitation here is the line rate. We are using 1 Gb/s Ethernet link and transmitting a collection of packets whose average size is 192 bytes. The maximum frame rate for the packets with frame size of 192 is around 500,000 frames per second~\cite{cisco}. 
Since we are sending one symbol per 256 packets, our maximum bandwidth is theoretically bounded at 1953 symbols per second. By coding three symbols, this packet chasing covert channel can reach a bandwidth of 3095 bps.
The error rate, however, is reduced as we reduce the probe time. That is because with a longer wait time between two consecutive probes, we raise the probability of capturing irrelevant background activity on the sets. When we use binary encoding, we use the samples from both \emph{set 2} and \emph{set 3} and if they both have activity during a window, we decode as "1". Therefore, the error rate is slightly less than the ternary encoding.   

\paragraph{Exploiting Ring Buffer Sequence Information}
If we know the ordering of the buffers, this mechanism is easily
extended to send more than one symbol per 256 packets.
In this case, the trojan can send one covert message every \textit{256/n} packets by dividing the ring buffer into \textit{n} sections of similar sizes by selecting \textit{n} buffers that are ideally \textit{256/n} apart. 
The selected buffers should be among the buffers that are mapped to only one of the page-aligned cache sets.  
Then the spy starts monitoring the selected sets and their adjacent blocks to detect the size of the packets that are filling these buffers. 

\begin{figure}[!t]
  	\centering
  \begin{subfigure}{0.24\linewidth}
  	\centering
  	\includegraphics[width=\textwidth]{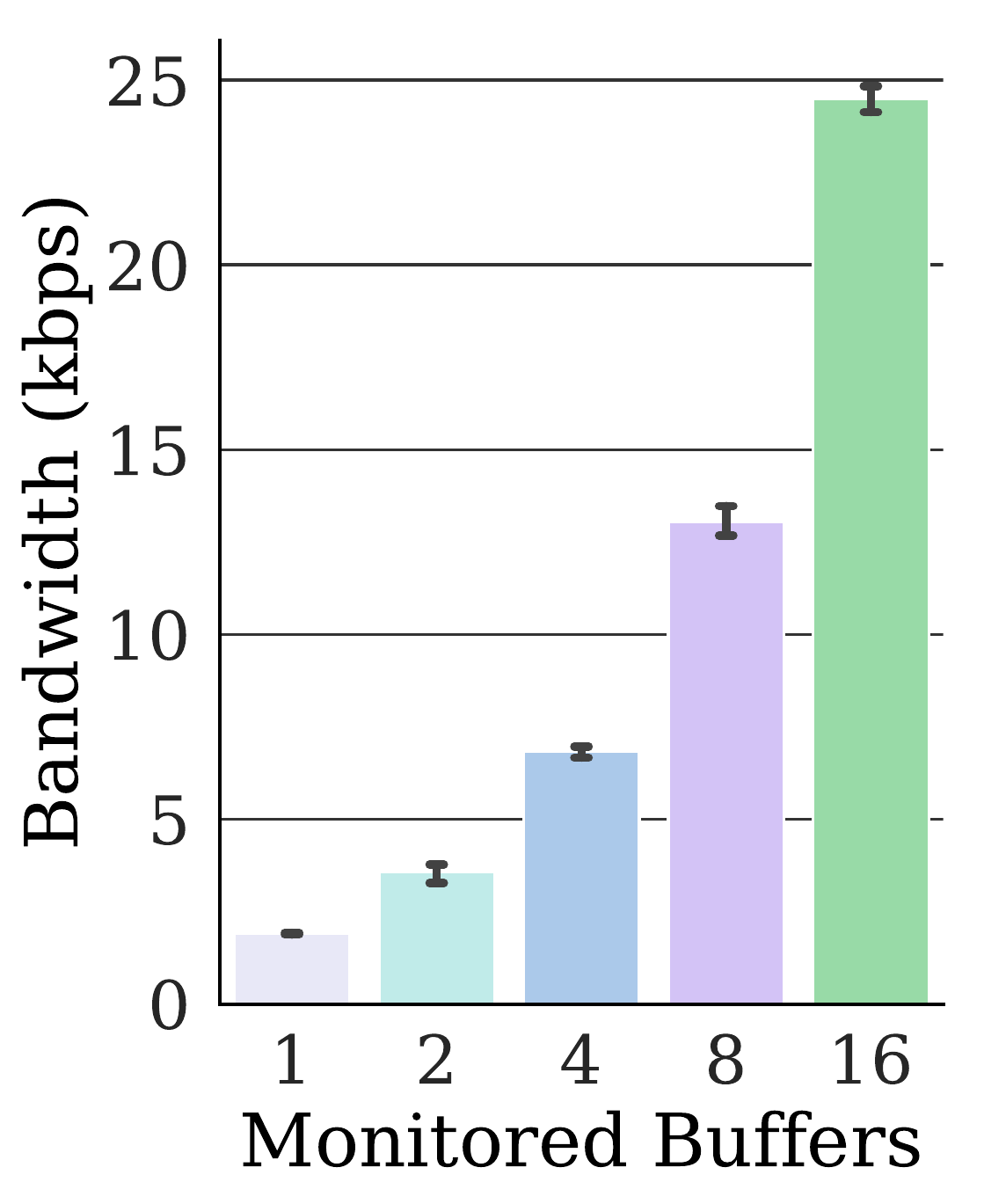} 
  	\vspace{-3ex}
    \caption{Bandwidth }
    \label{fig:covert_seq:bw}
  \end{subfigure}
  \begin{subfigure}{0.24\linewidth}
  	\centering
  	\includegraphics[width=\textwidth]{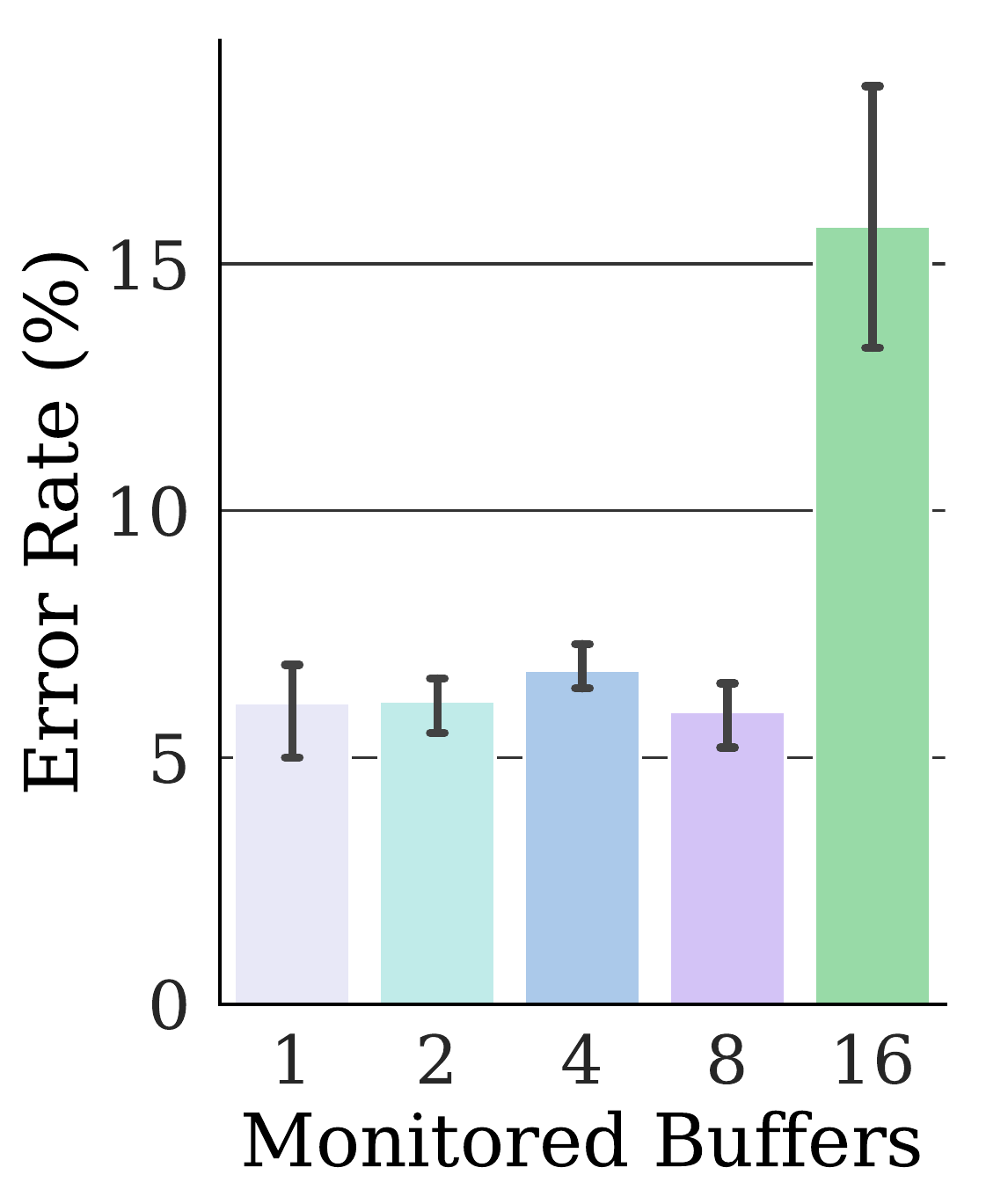} 
  	\vspace{-3ex}
    \caption{Error Rate}
    \label{fig:covert_seq:error}
  \end{subfigure}
   \begin{subfigure}{0.24\linewidth}
  	\centering
  	\includegraphics[width=\textwidth]{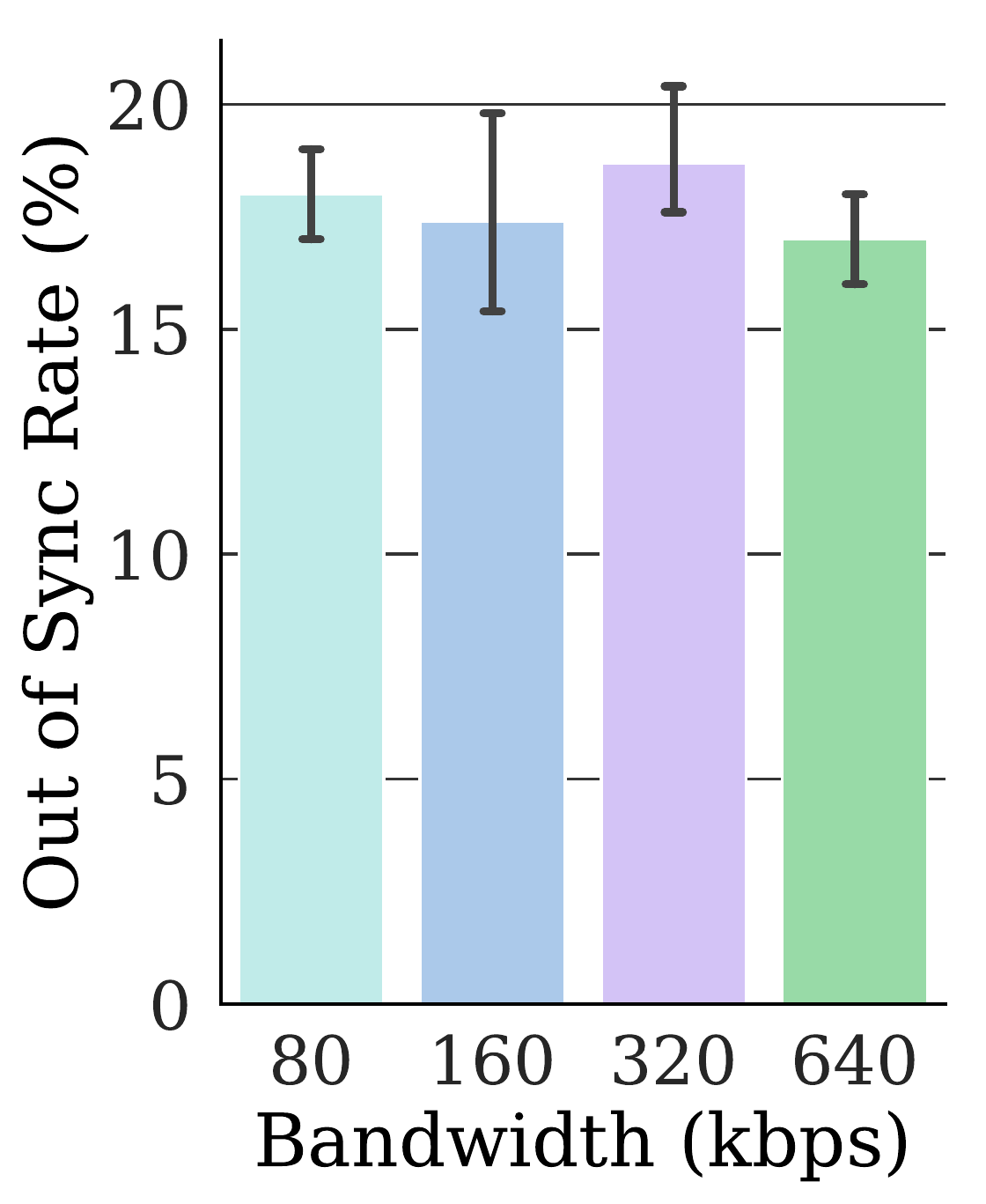} 
    \caption{Out of Sync}
    \label{fig:covert_chase:out}
  \end{subfigure}
  \begin{subfigure}{0.24\linewidth}
  	\centering
  	\includegraphics[width=\textwidth]{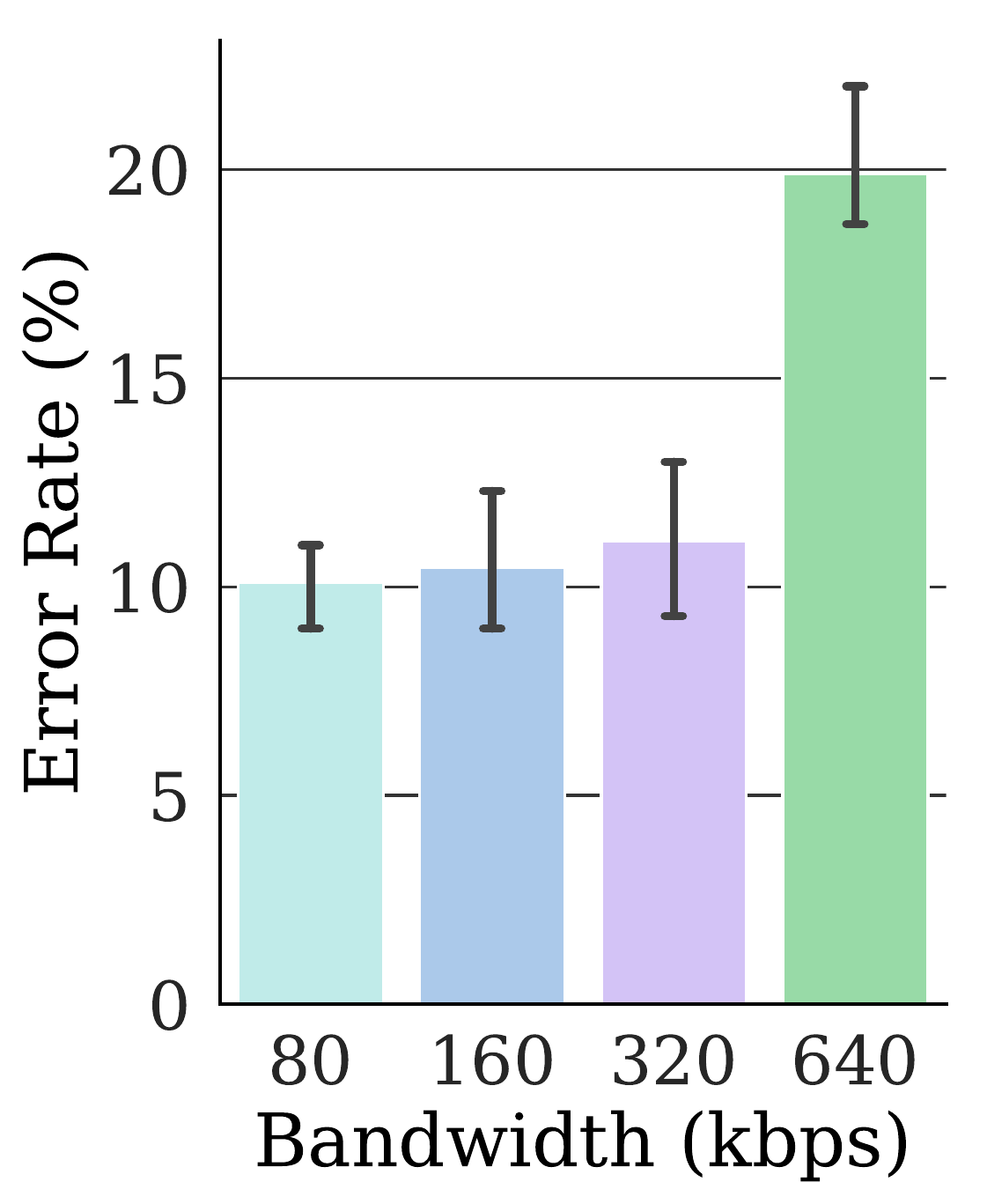} 
    \caption{Error Rate}
    \label{fig:covert_chase:error}
  \end{subfigure}
    \vspace{-.5ex}
  \caption{ (a) and (b) show the channel capacity for the remote covert channel where the spy that uses n buffers of the ring's sequence information. (c) and (d) show the out of sync rate and error rate for the case where spy uses all the buffers in the sequence information.}
\end{figure}

This process can multiply the capacity of the covert channel as shown in Figures~\ref{fig:covert_seq:bw} and~\ref{fig:covert_seq:error}.
These figures show the bandwidth and the error rate for the cases that the spy monitors a different number of buffers in the ring. 
For each of these buffers the spy probes three cache sets that are associated with the first, third, and fourth cache blocks of the packets that fill these buffers. 
For the case that there is only one monitored buffer, the trojan sends one covert message with 256 packets, and for the case of 16, the trojan sends a new message every $256/16=16$ packets.
The bandwidth of the channel almost doubles when we double the number of monitored buffers and it goes up to 24.5 kbps for the case of 16 monitored buffers. 
The error rate remains almost constant until the time between incoming packets gets close to the time between two consecutive probes. 
Note that when we have more sets in our monitored list, each probe takes more time and this decreases the probe rate. 
Furthermore, with the increased number of monitored buffers, it becomes harder to find the buffers that are \textit{n} buffers apart in the ring and also do not share the cache set with any other buffer in the ring. 
For these reasons, we see a jump in the error rate when we monitor 16 buffers of the ring. 
Note that these and subsequent results also account for inaccuracies incurred when we deconstruct the ring sequence.

Figures~\ref{fig:covert_chase:out} and ~\ref{fig:covert_chase:error} show the result of another experiment in which we actually chase the packets using the sequence. We probe one buffer at a time and as soon as we detect an activity on the probed buffer, we move to the next buffer in the sequence. 
The \emph{out-of-sync} rate is the rate by which packet chasing misses one packet, and therefore it has to wait until completion of the whole ring, or the next time a packet fills that buffer, to get synchronized again.  
The bandwidth is controlled by the rate at which the sender transmits the packets and the error rate is calculated on the synchronized regions of the transmission. 
The figure shows that the out-of-sync rate is almost constant for different packet rates. This is because when we probe just one set, the resolution of probing is higher than the time between consecutive packets. In addition, the frequency at which we get out-of-sync is a function of the quality of the sequence that we obtain. 
The error rate jumps at 640 kbps because at that speed the packets start to arrive out-of-order at the receive side.

\paragraph{Detectability and Role of DDIO/DCA }
\hl{In the presence of DDIO,} the packets that carry the covert messages are hard to detect and filter (\hl{e.g., by a firewall system that drops the packets that are sent to the victim node}) as they can be regular broadcast packets, e.g., DHCP and ARP, and they are not even required to be destined for the machine that hosts the spy. This is because, with DDIO/DCA, the network adapter directly transfers the packets into the last level cache of the processor, and only after this will the driver examine the header of each frame and discard the packets that do not target any protocol that is hosted in that machine. That is, with DDIO enabled, we can establish a channel between machine A and B
even with A only sending packets to machine C on the same network. 

\hl{DDIO enables Packet Chasing to get a clearer signal as the cache blocks of the payload appear in the cache as the same time as the cache blocks that belong to the header of the packet.
This enables the attack to probe the adjacent cache blocks and quickly detect the size of each packet. 
However, if DDIO is disabled or not present, the journey of a packet would be different. First, the NIC stores the header of the packet in the memory, then the driver reads the header and processes the packet according to the header fields. 
This brings the cache blocks containing the header into the cache.
For most of the common higher level protocols (e.g., http) the software stack will access other parts of the packet shortly after the header comes~\cite{dca}. }

\hl{
The latency between I/O writes and driver reads now becomes a factor in the attack without DDIO. In this scenario, the attacker should set the probe time to be larger than that latency.
When that latency is accounted for, the cache footprint of the packet remains the same as the DDIO case.
However, increasing the probe interval can result in more noise captured in each interval.
But the latency, as characterized in~{\cite{dca}}, is less than 20k cycles for almost 100\% of the packets.
%
This latency also depends on the size of the packets. For small packets with less than 5 cache blocks, the payload will be touched almost immediately after the header, as the driver copies such packets into another buffer. In this case, the attack without DDIO detects packet sizes for the small packets as readily as the attack with DDIO.}

\hl{In short, DDIO makes the attack stealthier, and more reliable (less noise). But the attack is fully possible without DDIO. As an example, the web fingerprinting attack presented in the next section is mounted on a system both with and without DDIO. 
}

\section{Packet Chasing: Exploiting Packet Size}\label{sec:side}
\begin{figure*}[!t]
  	\centering
    \begin{subfigure}{0.245\textwidth}
  	\centering
  	\includegraphics[width=\textwidth]{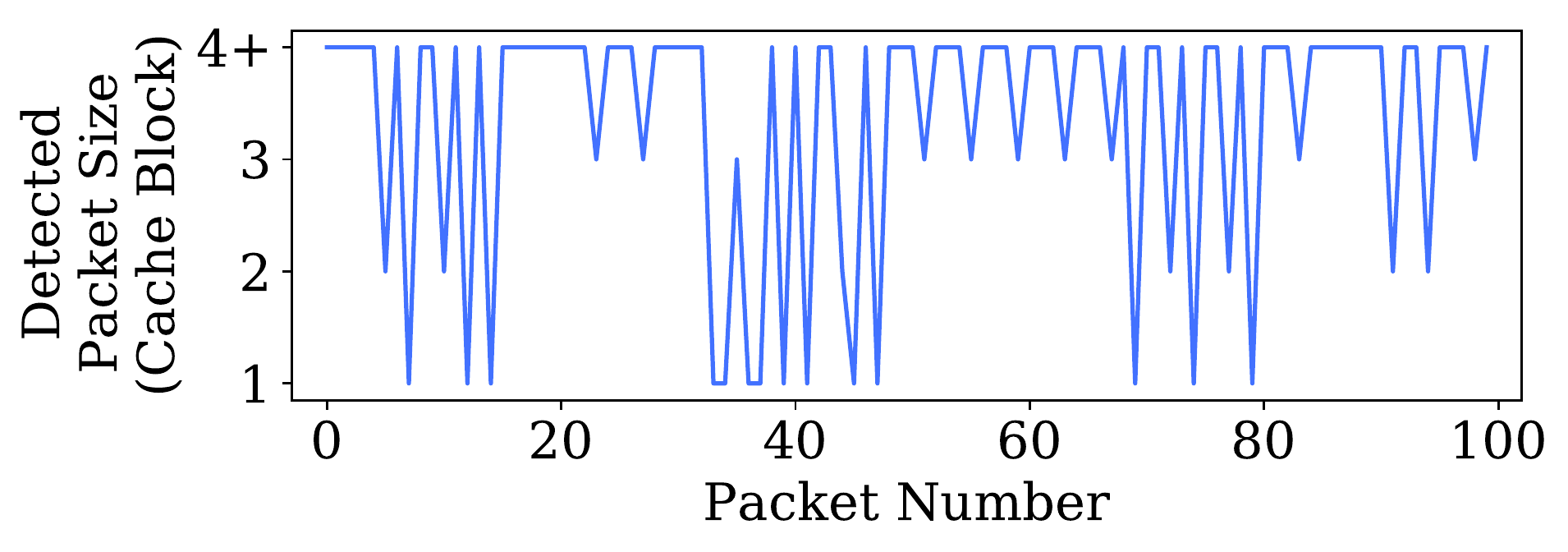} 
  	     \vspace{-4ex}
    \caption{Successful Login}
      
  \end{subfigure} 
  \begin{subfigure}{0.245\textwidth}
  	\centering
  	\includegraphics[width=\textwidth]{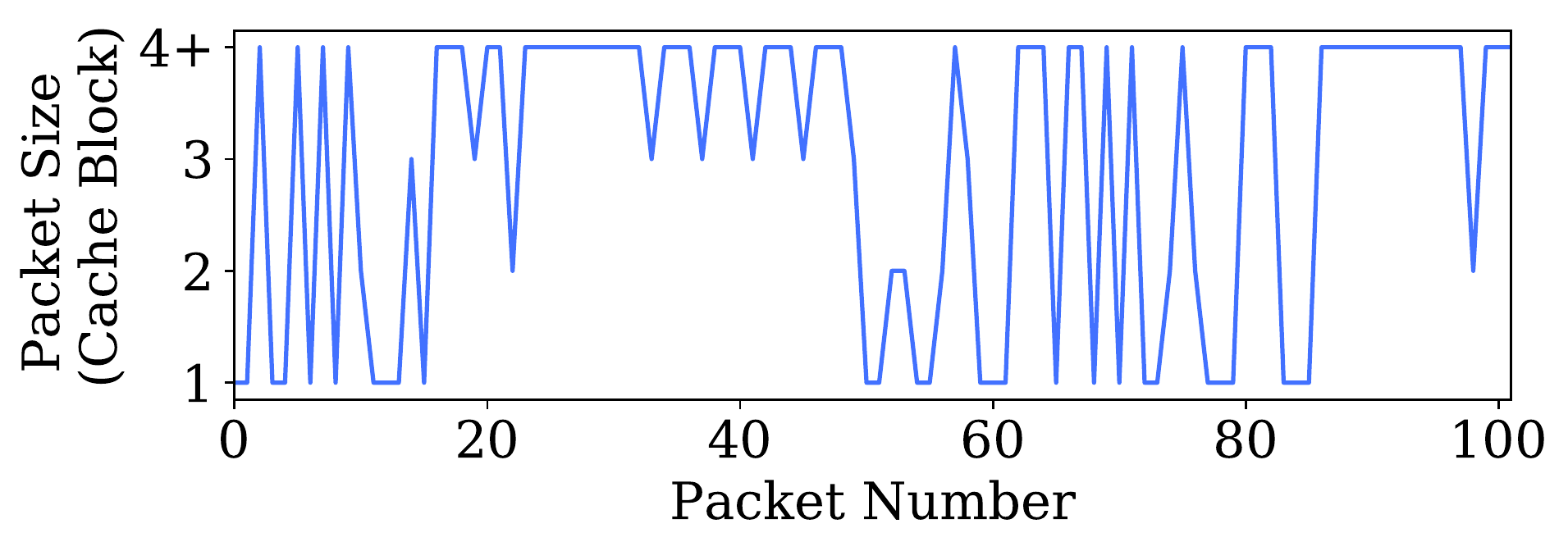} 
  	     \vspace{-4ex}
    \caption{Unsuccessful Login}
  \end{subfigure} 
   \begin{subfigure}{0.245\textwidth}
  	\centering
  	\includegraphics[width=\textwidth]{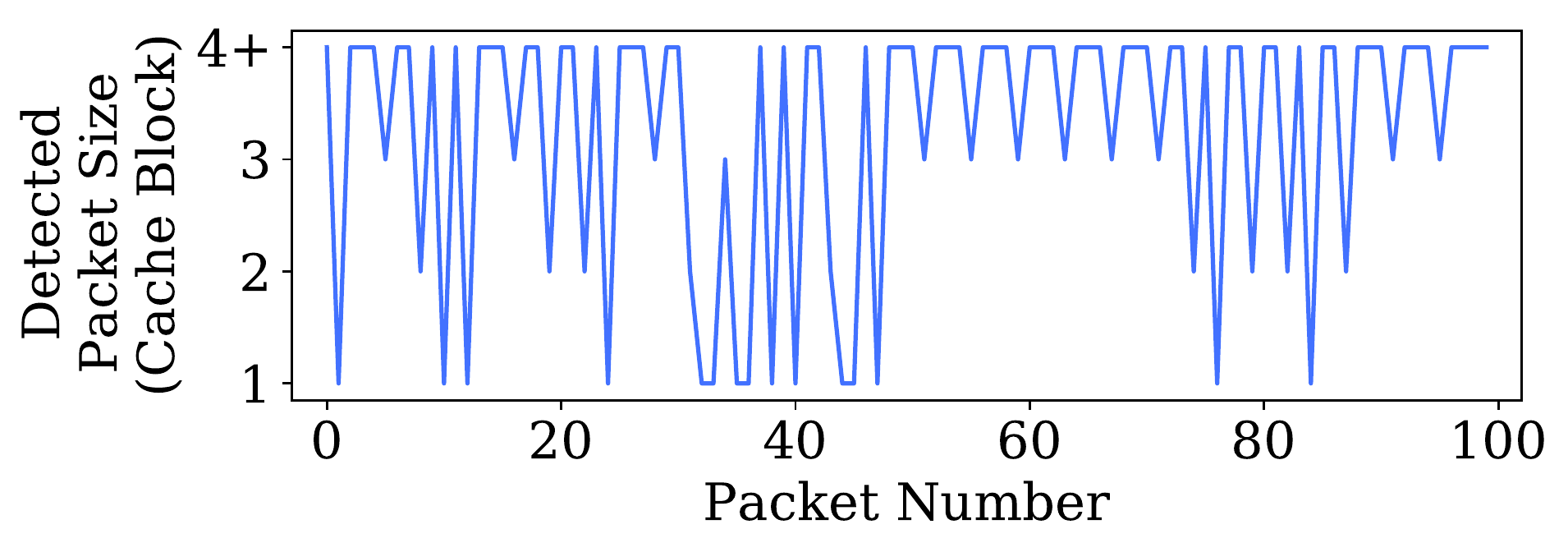} 
  	     \vspace{-4ex}
    \caption{Recovered Successful Login}
  \end{subfigure}
  \begin{subfigure}{0.245\textwidth}
  	\centering
  	\includegraphics[width=\textwidth]{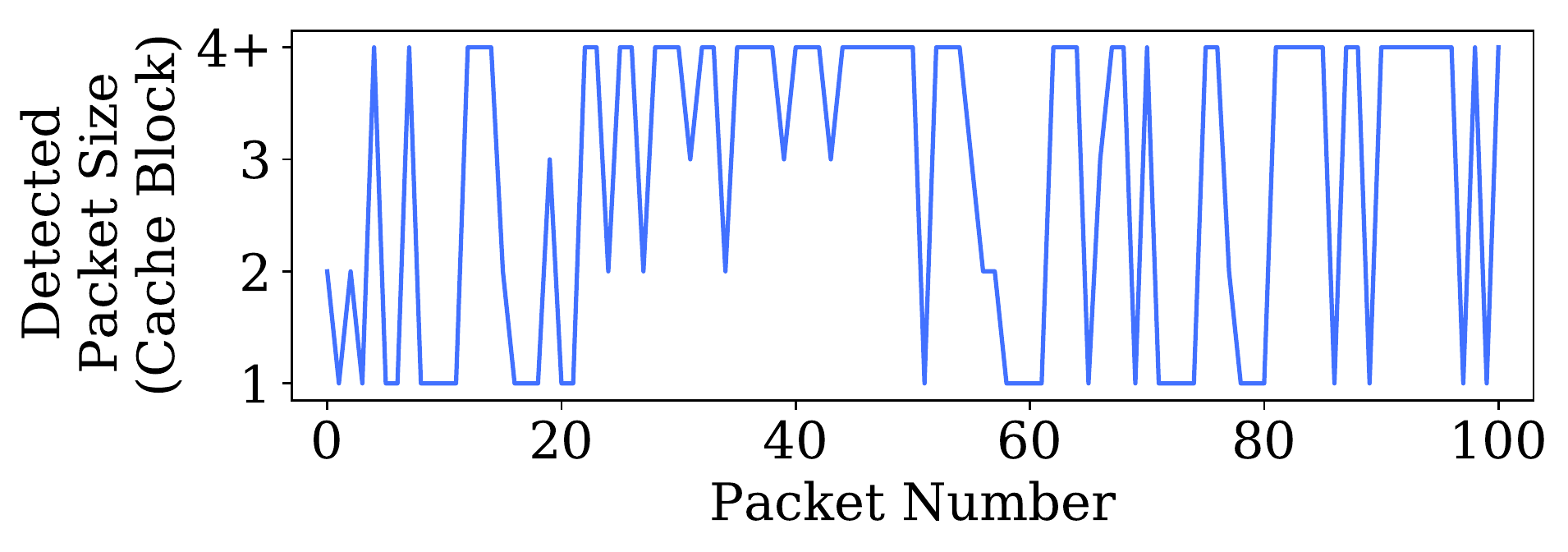} 
  	   \vspace{-4ex}
    \caption{Recovered Unsuccessful Login}
  \end{subfigure}
  
  \caption{Detecting successful login for hotcrp.com.  Shows original packet sizes vs. the recovered packet sizes by Packet Chasing for the first 100 packets of the responses.}
  \label{fig:side-channel}
\end{figure*}

In this section, we present a sample application for a Packet Chasing attack, in which we use the high resolution samples of packet sizes to gain information on the co-located user's browsing data. For example, the spy could be waiting for the victim to enter a particular website before initiating some action
such as a password detection attack.

This simple attack consist of two phases. First is the offline phase in which the adversary generates traces of packet sizes for different websites
of interest, then processes these traces and calculates a representative trace for each website. This is just a point-wise average of the packet sizes, 
resulting in a vector of these points (average packet size) over time.

In preparation for the attack, the attacker builds the sequence of the ring buffers, as previously described. 
After that, the attacker enables spy mode in which she constantly monitors the first two cache blocks of the first buffer in the sequence until she finds a window in which there are activities on both block 0 and block 1. This indicates that a packet is filling that buffer. %
Then, similar to the receiver in the covert channel, on each detected activity the attacker moves to the next buffer in the sequence. Each time, the attacker monitors the first four blocks of the first half-page of the buffer as well as the first four blocks of the the second half-page of the buffer. This is because the driver switches between the halves of the pages when there is a large packet (see Section~\ref{sec:deconstruction}).  This enables the attacker to distinguish between packets with four level of sizes.  
After collecting the samples of packet sizes, the spy feeds the collected vector into a simple correlation-based classifier which calculates cross-correlation~\cite{cross} of the collected samples with the representatives of different targets. 

Figure~\ref{fig:side-channel} shows an example of the signals that we obtain by Packet Chasing and the actual packet sizes that are captured using the tcpdump~\cite{tcpdump} packet analyzer. 
The websites are accessed using Mozilla Firefox version 68.0.1.
The figure shows how packet size, even in cache block granularity, can be an identifier for the webpages that are being accessed. 
The packets are usually congested on the two sides of the spectrum, they are either carrying a very large message that got fragmented into MTU-sized frames, or they are small control packets~\cite{Sinha07a}.  
But the last packet of the large messages can fall anywhere between 1-block to MTU, giving us a good indicator of the webpages. 
In addition, combining packet sizes with the temporal information that Packet Chasing obtains from the packets, gives us enough information to distinguish between webpages.
We evaluate our fingerprinting attack using a small closed world dataset with 5 different webpages: facebook.com, twitter.com, google.com,  amazon.com, apple.com. 
\hl{For this experiment, we examined two attack setups, one with DDIO and one 
without.
In our experimentation with 1000 trials, Packet Chasing with DDIO detects the correct website with 89.7\%  accuracy, while disabling DDIO drops the accuracy to 86.5\%.
The difference between these two attacks comes from increasing the probe time
(resulting in more noise) and increased probability of missing large packets if the
header-to-payload latency is high.}

We use a simple classifier in this experiment, but given the challenges for this particular attack,
a classifier that is tolerant of noise as well as slight compression or decompression of the vectors would be likely to improve these results. { For example, the results in~\cite{Rimmer2018} suggests that using only the network packet sizes and their timing information (the exact information that Packet Chasing can obtain) can be enough to build a classifier with up to 95\% accuracy. }

\section{Potential Software Mitigation}\label{sec:potential:soft}
\begin{table}
  \caption{Architecture Detail for Baseline Processor}
  \label{tab:gem5-arch}
  \vspace{-1em}
  \scriptsize
  \begin{center}
    \begin{tabular}{|l|l||l|l|}
      \hline
      \multicolumn{4}{|c|}{\textbf{Baseline Processor}}                      \\
      \hline
      Frequency & 3.3 GHz                           & Icache & 32 KB, 8 way       \\
      Fetch width & 4 fused uops               & Dcache & 32 KB, 8 way       \\
      Issue width & 6 unfused uops               & ROB size & 168 entries         \\
      INT/FP Regfile & 160/144 regs             & IQ & 54 entries \\
            RAS size & 8,16, 32 entries                         & BTB size & 256 entries\\
      LQ/SQ size & 64/36 entries                    & Functional & Int ALU(6), Mult(1)  \\
        
      \hline
    \end{tabular}
  \end{center}
\end{table}

We consider both long-term (e.g., requiring hardware changes) and short-term (software only)
mitigations. 
In this section, we discuss potential software mechanisms that one could employ to help mitigate the attack before a long-term hardware solution (e.g., our I/O cache isolation) is deployed.
These solution each come with some performance impact.

\paragraph{Disabling DDIO/DCA}
DDIO enables these attacks because it ensures the header and the payload appear in the
cache simultaneously, greatly simplifying the detection of packet size.  Without this, however, attacks are still
possible.  If we can detect the presence of packets (headers are always accessed immediately and will appear in the
cache in sequence), we can still establish a covert channel with inter-arrival timing.  We could also send types
of packets
where the reading/processing of the payload is quick and deterministic, again allowing us to distinguish sizes. Therefore, disabling DDIO can not fully mitigate the attack.

\paragraph{Randomizing the Buffers}
While Packet Chasing exploits the sequence in which the packets fill the ring buffers to boost the resolution of the side- and covert-channels, we show that attacks are still possible, without knowing the sequence of the buffers (Section~\ref{sec:covert}). 
However, randomization does significantly reduce the channel bandwidth.  The cost of randomization could be quite high, as the driver and the network adapter would now need to constantly synchronize on the address of the next buffer.
Because our attack setup takes some time, though, it may only be necessary to permute the buffer order at semi-regular
intervals, thus limiting the overhead.

\paragraph{Increasing the Size of the Ring}
In the absence of sequence information, the required probing of the cache scales with the size of the ring if the
attacker wants to catch every packet.  Thus a combination of occasional reshuffling of the ring, and a larger ring,
may be effective in making the probe set large enough to make the attack difficult to mount cleanly without picking
up significant noise.

\section{Adaptive I/O Cache Partitioning Defense}
All the short-term software mitigations that Section~\ref{sec:potential:soft} suggests are either not fully effective (disabling DDIO), or carry a not-insignificant performance cost.
In this section, we describe a hardware defense that tackles the root of the vulnerability, i.e., co-location of I/O and CPU blocks in the last-level cache, in such a way that I/O can cause evictions
of other processes' lines.

Intel's DDIO technology improves the memory traffic by introducing a last-level cache write allocation for the I/O stream.
Upon receiving a write request from an I/O device (e.g., for incoming packets), DDIO allocates cache blocks in the last-level cache and sets those blocks as the DMA destinations for the incoming I/O traffic. 
While for performance reasons, the allocator does not allocate more than two blocks in a cache set, these incoming packets can still cause evictions to the CPU's blocks. 
This makes the incoming packets observable from the perspective of an adversary process running on the CPU. 

To circumvent this, we associate a counter for each set ($i$) to hold the size of the I/O partition ($IO\_lines_i$). By treating this as a constant (during a single interval) rather than a maximum, we ensure that DDIO-filled lines will only 
displace other DDIO lines.
To adapt to different phases of execution, our partition schemes periodically change the boundary of CPU and I/O partitions by incrementing or decrementing the counter ($IO\_lines_i$). 
To this end, we associate another set of counters for each set to detect the I/O activity on each set ($IO\_present\_counter_i$). This counter gets incremented if at least one valid I/O line is present in the set, and initialized to zero at every adaptation period cycles ($p$). 
Note that maintaining these counters does not impose a performance overhead as these are done in parallel to the miss and hit path of the cache. 

At every adaptation period, we also re-evaluate the I/O-CPU boundary in the last-level cache. For each set ($i$), if the ($IO\_present\_counter_i$) is greater than a high threshold, $T_{high}$ (e.g, $T_{high}=0.5p$), it implies $set_i$ has had significant I/O activity. In such a case, we increment $IO\_lines_i$ (using a saturating counter), allowing more I/O blocks in the set. Otherwise if the I/O activity is less than a low threshold, $T_{low}$,  we decrement $IO\_lines_i$ (again, using a saturating counter) to allow more usage for CPU data. If the boundary of the partitions changes, we invalidate the cache blocks that are affected and perform any necessary writebacks to the memory. 

{Our adaptive partitioning ensures that any process running on the CPU will not see any of its cache lines evicted as the result of an incoming packet or I/O activity.
The only exceptional scenario is at each adaptation period when the boundary changes and some CPU blocks get evicted. However, we set the adaptation period to be large enough to prevent the attacker from gleaning any useful information about individual packets.  At best, it could receive one bit (high or low 
network activity) every period.} 

\paragraph{System Setup for the Defense Performance Evaluation}
 Table~\ref{tab:gem5-arch} shows the architectural configuration of our baseline processor in detail. We model this architecture using the gem5~\cite{binkert06} architectural simulator. 
 We use the full system simulation mode of gem5 which allows us to boot an Ubuntu 18.04 distribution of Linux with a kernel version of 4.8.13. 
 We set a hard limit on the minimum and maximum number of blocks in the I/O partition (i.e., $IO\_lines_i$). As such, the size of the I/O partition can be one, two, or three. Also, in these experiments the adaptation period ($p$) is set to 10k cycles. We set the thresholds $T_{low}$ and $T_{high}$ to 2k and 5k, respectively.   
 Furthermore, in order to provide realistic estimates regarding the performance impact of our proposed defense, we select a mix of benchmarks that exhibit considerable amounts of I/O activity.  To this end, we include a disk copy (using Linux's \emph{dd} tool) that copies a 100MB file from disk. In addition, we evaluate the defense on a program that constantly receives TCP packets that have 8-byte payloads. Finally, we also evaluate the impact of our defense on the \emph{Nginx} web server~\cite{nginx} using the \emph{wrk2}~\cite{wrk} framework to generate HTTP requests.

\begin{figure}
    \centering
    \includegraphics[width=0.5\textwidth]{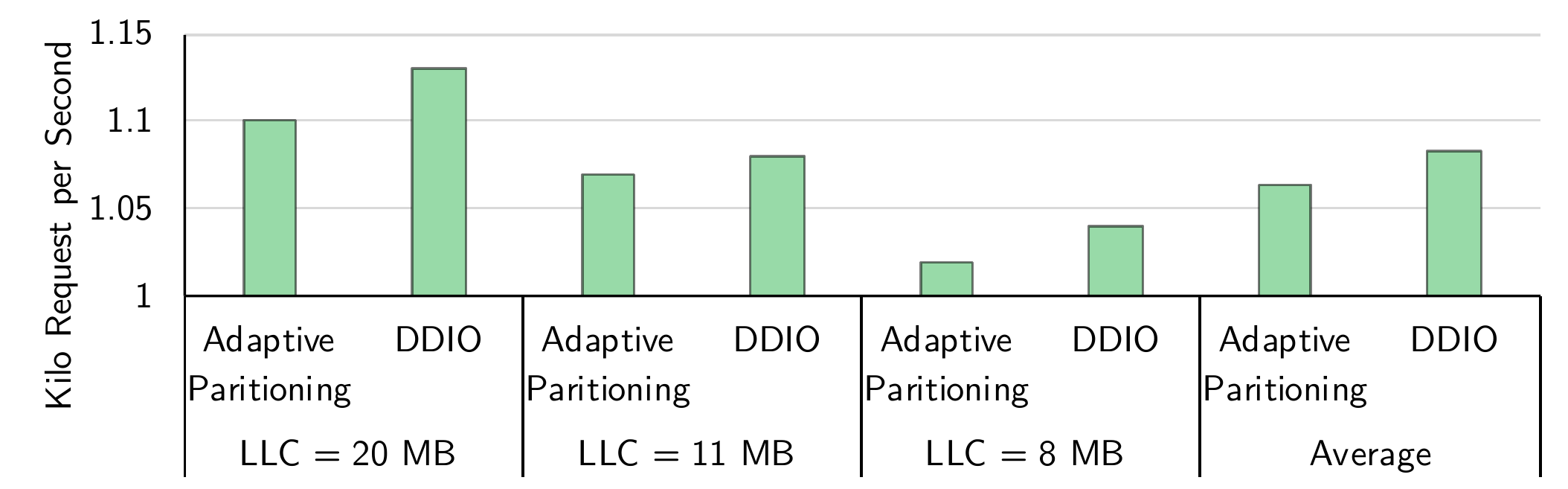}
    \caption{Performance impact of our adaptive partitioning defense on Nginx web server.}
    \label{fig:defense:results1}
\end{figure}

\begin{figure}
    \centering
    \includegraphics[width=0.5\textwidth]{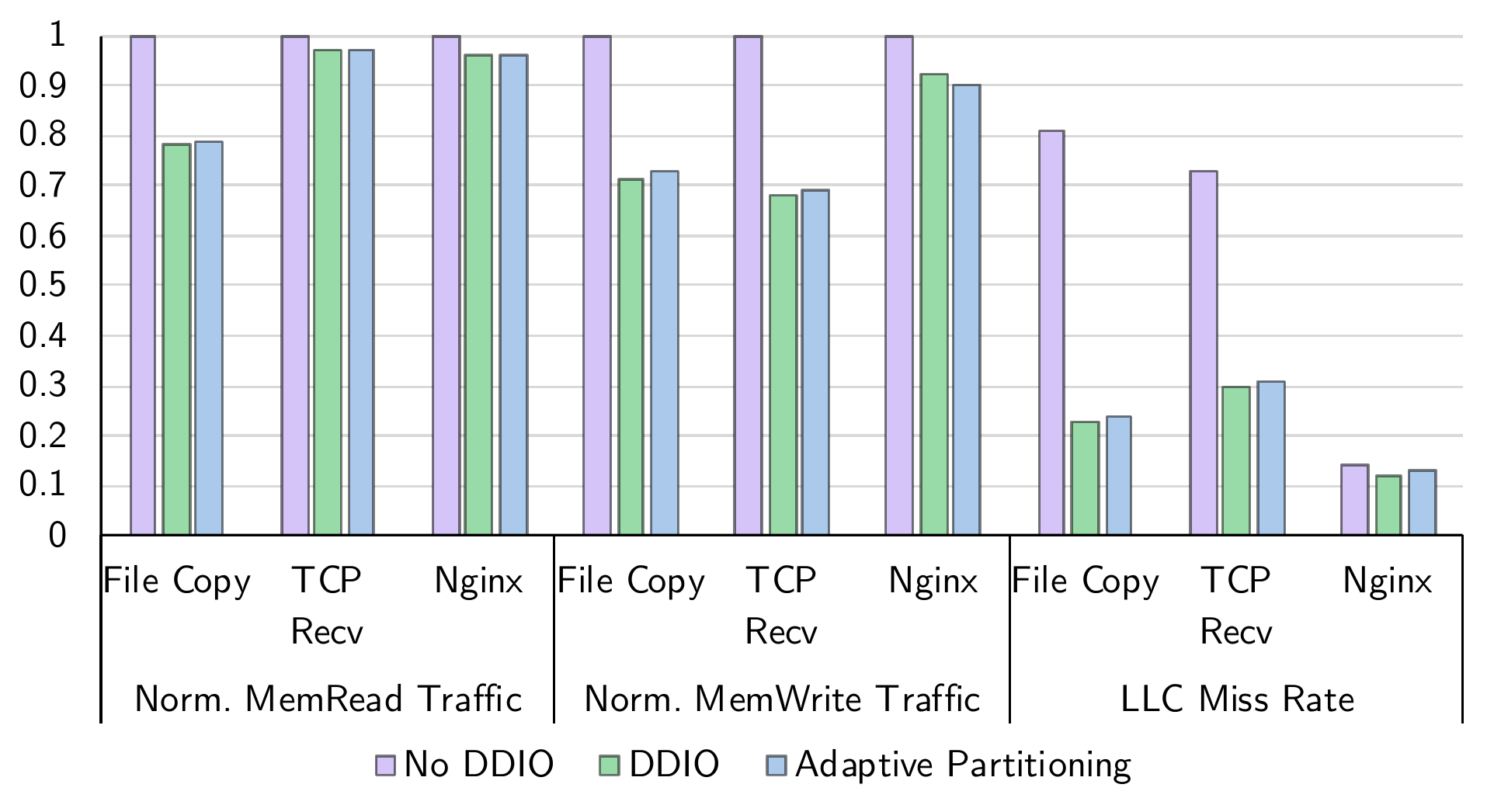}
    \caption{Memory Traffic and LLC miss rate of our adaptive partitioning defense vs. DDIO.}
    \label{fig:defense:results}
\end{figure}

\paragraph{Performance Results of the Defense}
Figure~\ref{fig:defense:results1} shows the performance of our adaptive cache partitioning scheme by comparing the average throughput of the Nginx web server. On average, we observe less than two percent loss in throughput. This is mainly because the LLC miss rate rises slightly due to the reduced number of lines in the CPU partitions (also see Figure~\ref{fig:defense:results}). 
The figure also shows the sensitivity of the defense to the last-level cache size. The maximum loss in throughput belongs to the 20 MB case where our approach incurs 2.7\% loss. 
Figure~\ref{fig:defense:results} further analyzes the performance of the defense by showing the memory traffic and the LLC miss rate of a baseline without any direct cache access (\emph{No DDIO}) vs. DDIO and our adaptive partitioning defense. Both the adaptive partitioning and DDIO are effective in reducing the memory traffic. The memory traffic of the adaptive partitioning scheme is within 2\% of DDIO. 

\begin{figure}
    \centering
    \includegraphics[width=0.95\linewidth]{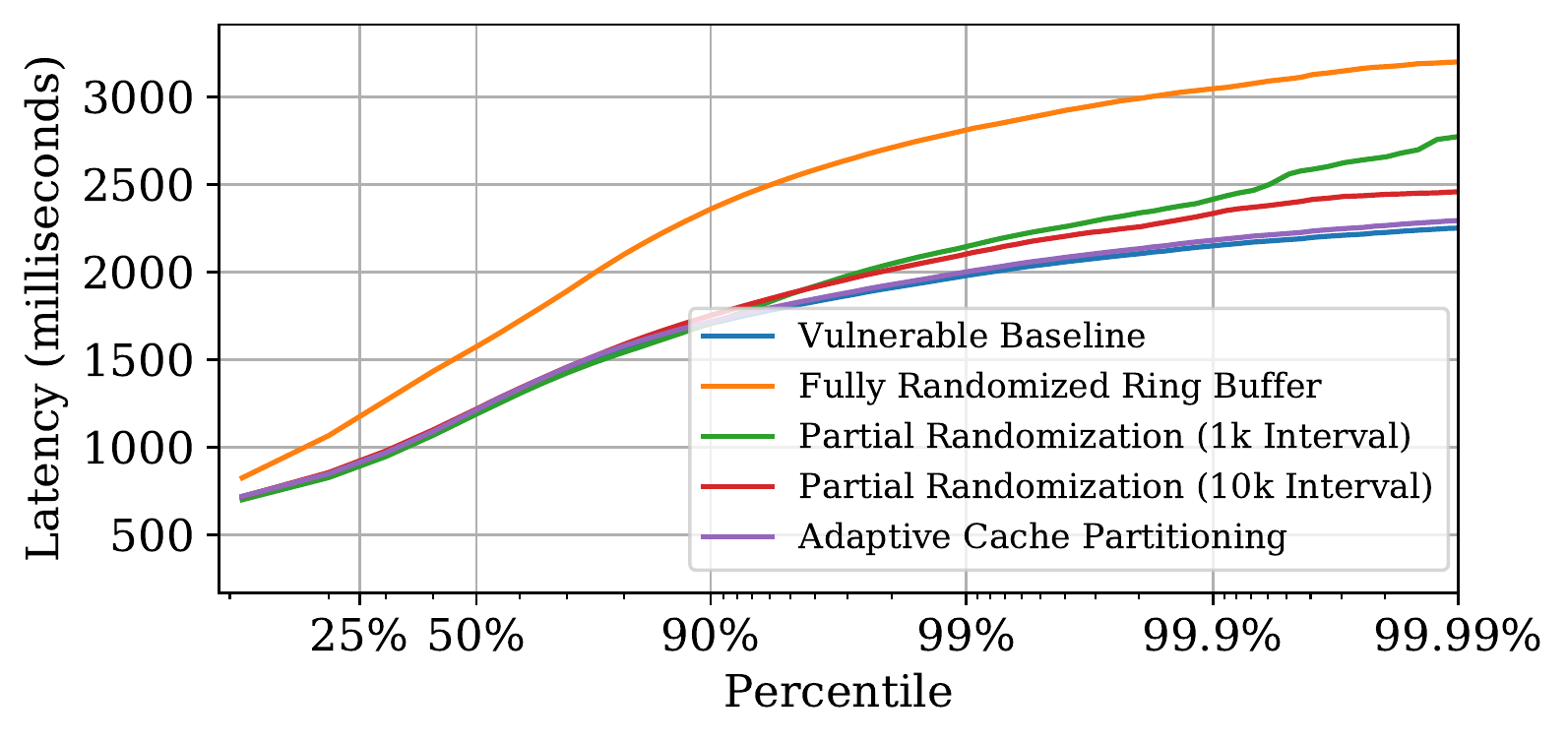}
    \caption{\hl{Comparison of our defenses in terms of response (tail) latency of HTTP requests to the Nginx web server. Randomization period is the interval (measured in number of packets) that we wait between two ring buffer randomizations.}}
    \label{fig:defense:tail}
    \vspace{1ex}
\end{figure}

\hl{To compare the adaptive cache partitioning with our proposed software-based mitigations (Section~\ref{sec:potential:soft}), we devise another experiment using the wrk2 tool. In this experiment we send requests to the Nginx web server on the target host.
The wrk2 tool uses eight threads with 1000 open connections and the target throughput is set to 140k requests per second. 
Figure~\ref{fig:defense:tail} shows the results of this experiment.  Besides the adaptive cache partitioning and the vanilla IGB baseline, we examine three other proposed schemes:  \textit{Fully Randomized Ring Buffer} scheme that allocates a new buffer in a random memory location for each incoming packet, and two \textit{Partial Randomization} schemes that re-allocate the buffers periodically, after a specified number of packets received -- we randomize after either 1k or 10k packets 
are received.
Note that in our setup, the Packet Chasing attack currently requires at least 65,536 packets to fully deconstruct the ring buffers (find cache locations and sequence information) and another 100 packets to mount a reasonable fingerprinting attack. 
The adaptive partitioning method only incurs 3.1\% loss in 99\textsuperscript{th} percentile latency while the fully randomized method incurs 41.8\%. We use 
One Gigabit Ethernet for this experiment, but we expect the performance cost of randomization to be exacerbated as the link rate goes higher. }
\section{Disclosure}

{We disclosed this vulnerability to Intel, explaining the basic substance of the vulnerability and offering more details. MITRE has assigned an entry in the Common Vulnerabilities and Exposures (CVE) database, CVE-2019-11184.  The vulnerability is classified as a medium severity vulnerability.}
\section{Conclusions}
This paper presents Packet Chasing, a novel deployment of cache side-channel attacks that detects the frequency and size of packets sent over the network,
by a spy process that has no access to the network, the kernel, or the process(es) receiving the packets.  This attack
is not enabled by the DDIO network optimization, but is greatly facilitated by it.  This work shows
that the inner workings of the network driver are easily deconstructed by the spy process setting up the attack, 
including the exact location (in the cache) of each buffer used to receive the packets as well as the order in which
they are accessed.  These two pieces of information dramatically reduce the amount of probing the spy must do to follow the network packet sequence.
This information enables several  covert channels between a remote sender and a spy anywhere on the network,
with varying bandwidth and accuracy tradeoffs.  It also enables a side channel leakage attack that detects
the web activity of a victim process.

{In addition to the covert- and side-channel attacks, this paper also describes an adaptive cache partitioning scheme that mitigates the attack with very low performance overhead compared to the vulnerable DDIO baseline.}
\section*{Acknowledgment}

The authors would like to thank the anonymous reviewers for their helpful insights. This research was supported in part by NSF Grants CNS-1652925 and CNS-1850436, NSF/Intel Foundational Microarchitecture Research Grants CCF-1823444 and CCF-1912608, and DARPA under the agreement number HR0011-18-C-0020.

\bibliographystyle{ieeetr}
\bibliography{references}

\end{document}